\documentclass[aip,pop,sd,reprint,10pt,amsmath,amssymb,graphicx]{revtex4-1}

\usepackage[capitalize]{cleveref}
\usepackage[T1]{fontenc}
\usepackage{float}
\usepackage{rotating}
\usepackage[normalem]{ulem}
\usepackage{amsmath}
\usepackage{textcomp}
\usepackage{color}
\usepackage{marvosym}
\usepackage{wasysym}
\usepackage{amssymb}
\usepackage{bm}

\newcommand{\beq}{\begin{equation}}
\newcommand{\eeq}{\end{equation}}
\newcommand{\bfig}{\begin{figure}[htbp]}
\newcommand{\efig}{\end{figure}}

\newcommand{\ben}{\begin{eqnarray}}
\newcommand{\een}{\end{eqnarray}}
\usepackage{graphicx}
\usepackage{color}
\usepackage{subfig}
\usepackage{subfig}
\usepackage{epsfig}
\usepackage[titletoc,toc,title]{appendix}

\begin{document}
\title{Coherent  transport structures  in magnetized plasmas  {\it II} : \\ Numerical results}
\author{G. Di Giannatale}\affiliation{IGI - CNR Corso Stati Uniti 4, Padova, Italy}
\author{M.V. Falessi}\affiliation{ENEA C.R. Frascati Via Enrico Fermi 45  Frascati, \\ Dipartimento di Matematica e Fisica,  Roma Tre University,  Roma, Italy}
\author{D. Grasso}\affiliation{ISC - CNR and Politecnico di Torino\\  Dipartimento Energia C.so Duca degli Abruzzi 24, Torino. Italy }
\author{F. Pegoraro}\affiliation{Dipartimento di Fisica E. Fermi,  Pisa University, \\ largo Pontecorvo 3, Pisa, Italy}
\author{T.J. Schep}\affiliation{Fluid Dynamics Laboratory, Dep. Applied of Physics, Eindhoven University of Technology, p.o.box 513, 5600MB Eindhoven, The Netherlands}

\bigskip

\begin{abstract}

  In a pair of linked  articles (called Article {\it I} and  {\it II}  respectively)  we apply the concept of Lagrangian Coherent Structures borrowed from the study of Dynamical Systems to magnetic field configurations in order to separate regions where field lines have different kind of behavior.\\
 In  the present article, article {\it II},  by means of a numerical procedure we investigate  the Lagrangian Coherent Structures  in the case of a two-dimensional magnetic configuration   with   two island chains  that are  generated  by magnetic reconnection  and  evolve nonlinearly in time.
 The comparison with previous results, obtained by assuming a fixed magnetic field configuration, allows us to explore the dependence of transport barriers on  the particle velocity.
\end{abstract}

\maketitle

\section{Introduction}

{In recent years  the  concept  of Lagrangian
Coherent Structures (LCS) has been introduced  by G. Haller, i.e. see Ref.\onlinecite{haller2000lagrangian}, in the context of  transport processes in complex 
fluid flows.  In an accompanying paper,  referred to here as Article {\it I},   it was shown that  such a concept can be usefully applied to  the study of particle transport  in a magnetized plasma in the limit where the field line dynamics can be taken as a proxy for the particle dynamics. 
In particular in Article {\it I}   it was shown how to relate a magnetic  field configuration,  at a fixed physical time, to a Hamiltonian system  where the role of ``time''  (Hamiltonian time) is taken 
by an appropriately chosen coordinate along the magnetic field lines. In the same article, after a brief summary of  the so called ``lobe-dynamics'' and  of the related transport in a nonautonomous  one-degree of freedom Hamiltonian system, the definition and properties of the Lagrangian Coherent Structures (LCS) were recalled. In the case of  a  (Hamiltonian) time periodic configuration, i.e. of a configuration that is geometrically periodic in the direction of the magnetic field  as is the case e.g. of a toroidal configuration, the connection  with the widely used Poincar\'e map  approach   was  mentioned.  Finally in Article {\it I}  the magnetic configuration that  is used in the numerical simulations reported in the present paper was introduced and  a generalization to the case where the LCS 
are defined so as to include the evolution of the magnetic configuration in time was discussed.  The chosen magnetic configuration is based on the investigation presented in  Refs. {\onlinecite{borgogno2005aspects,borgogno2008stable,borgogno2011barriers}} of  the nonlinear evolution of two chains of magnetic islands produced by magnetic  reconnection.

In the present paper the concepts introduced in Article {\it I}   are implemented numerically using  {a MATLAB tool developed by K. Onu and G. Haller, see Ref.} {\onlinecite{onu2015lcs}}. First the LCS are obtained by considering a snapshot at a fixed  physical time of the evolving  magnetic configuration 
by exploiting explicitly its periodicity in  ``Hamiltonian''  time (see Eq.(20) of Article {\it I}) .   Then the same numerical procedure is used to include the evolution of the magnetic configuration in  physical time. This  allows us to explore the dependence of transport barriers on particle velocity.}

This paper is organized as follows.\,\, In Sec.\ref{SSP}, after recalling the main features of the magnetic configuration of interest,  we introduce the adopted numerical computation scheme   and briefly describe the precautions that have been used in its implementation.  
In Sec.\ref{resmagnper} we take the flux function in the magnetic   Hamiltonian at a fixed physical time: we choose  $t = 415$, i.e. before the onset of fully developed chaos. 
The  corresponding LCS  are  then obtained numerically  and compared  to the structures  in  the  Poincar\'e map: in fact  both methods can be used in this case since the corresponding dynamical system is periodic in Hamiltonian time.
In Sec.\ref{res}   we consider the case of a magnetic field that evolves in physical time, i.e.  the case 
where  a charged particle moving in the plasma sees a time varying magnetic field
during its motion, and apply the simplified model described in Sec. VIB of Article {\it I}. In this case the corresponding dynamical system turns out not to be periodic in time and the Poincar\'e map technique cannot be applied.
The  LCS  are  then obtained numerically  for different particle streaming velocity along field lines  with the aim of  finding 
how do the LCS change with physical time, how they differ  from those found at the fixed physical time $ t = 415$, 
and, in addition,whether and how  particle with different velocities  can cross  LCS calculated for a different particle velocity. Finally the conclusions are presented.

\section{Simulation settings and  numerical procedure} \label{SSP}

\subsection{Magnetic configuration}\label{magnconfig}

As anticipated in Sec. VI  of Article {\it I},    in the present article  we study the LCS in  a magnetic configuration of the form 
\beq\label{equil}
    {\bf B}_{eq} =  B_0{\bf e}_z + \nabla{\psi}(x, y, z, t)\times {\bf e}_z ,
\eeq
where $\psi(x, y, z, t)$  is the 
full magnetic flux function that includes the equilibrium and the time evolving perturbations. Periodicity is assumed in all three directions and the configuration is restricted to the domain 
$[-L_x,L_x]\times [-L_y,L_y]\times [-L_z ,L_z]$  with   $L_x = \pi, L_y = 2\pi, L_z =16\pi$.
\\
We recall that the expression of   $\psi(x, y, z, t)$  that   we use    has been obtained  by means of a numerical simulation  in Ref.{\onlinecite{borgogno2005aspects}} (see also  Refs.{\onlinecite{borgogno2008stable,borgogno2011barriers}})  by imposing  a ``double helicity'' perturbation $ \hat{\psi}_1(x, t) \cos{(k_{1y}y + k_{1z}z)} + \hat{\psi}_2(x, t) \cos{(k_{2y}y + k_{2z}z)}$,
{with  $k_{iy} = m_i \pi/L_y$ and $k_{iz} = n_i \pi/L_z$ where $ m_1 = m_2 = 1$ and $n_1 = 1$, $n_2 = 0 $},  on an  equilibrium of the form $\psi_{eq}(x)\propto \cos{(x)}$. \\
The field line equations are given by 
\begin{equation}  \label{psin1} \frac{d x}{d z} = - \frac{\partial  \psi}{\partial y} , \qquad  \frac{d y}{d z} = \frac{\partial  \psi}{\partial x} . \end{equation}
Perturbations with different ``helicities''   are required in order to make the  Hamiltonian system  described in Sec. II  of Article {\it I}  non integrable, i.e., to generate a chaotic magnetic configuration.
In the following analysis  we will focus on the magnetic configuration  at two different normalized (respect to the Alfv\'en time) physical  times, {i.e.} $t = 415$ and  $t = 425$, in which chaos, initially developed only  on a local scale (at $t=415$), starts to  spread on a global scale (at $t=425$).

 In order to minimize the computational effort,  we  simplify the   Hamiltonian 
 by imposing a threshold condition  on the amplitude  of the components of  the Fourier expansion  of  $\psi(x,y,z,t)$  along $x, y $ and $z$. The validity of this approximation has been tested  in Ref. \onlinecite{borgogno2008stable}.
The physical time evolution  of  $\psi(x,y,z,t)$ between   $t = 415$ and  $t = 425$  was found  in Ref. \onlinecite{borgogno2005aspects}  to be super-exponential  and is modeled  here by interpolating the coefficients of its  Fourier expansion according to a quadratic  exponential time law of the form
\beq
\exp{\gamma_{k_y,k_z}(t-t_1)^{2}} \quad \text{for} \quad t>t_1 =415,
\label{interplaw}
\eeq
where  we 
assume that the  coefficients $\gamma_{k_y,k_z}$
depend only on  the  mode numbers $k_y$ and $ k_z$.

\subsection{LCS computation scheme}\label{intmet}

In order to find the hyperbolic Lagrangian Coherent Structures, we use  a MATLAB tool developed by K. Onu and G. Haller,  see Ref. \onlinecite{onu2015lcs}. This {tool detects} the LCS on the basis of their characterization
as the most repelling or attractive material lines  advected within the fluid and relies on the definitions that we listed  in  Secs. V and VA of  Article {\it I}. \\
The key steps of the adopted  procedure  can be  summarized by the following operations:\\
1. Defining a velocity field\\
2. Computing the eigenvalues and eigenvectors of the Cauchy-Green  strain tensor\\
3. Filtering the data by setting the value  of some parameters requested by the numerical tool and aimed at locating the most important LCS to characterize the  system dynamics. Detail on this filtering procedure will be given below.\\
4. Finally detecting the LCS\\
The procedure 
starts with the integration of the  Hamilton equations,  Eq.(\ref{psin1}), for the  magnetic field lines.  
This enables us to calculate  the flow map 
$ {\boldsymbol \phi}_{z_0}^z( x_0,y_0)$  (defined in Eq.(4)   of  Article {\it I})  with    $z$ substituted for $t$ and then to compute the Cauchy-Green strain tensor field, its eigenvalues and eigenvectors and the FTLE field.
The repelling LCS are then found following  the conditions given in Sec. V A of  Article {\it I}. In particular, in lieu of Eq.(14) %\ref{maxrep}
 but following Ref. \onlinecite{onu2015lcs}, we identify   the strongest repelling curves as those passing through a local maximum of the FTLE field.

The advantage of such a  prescription is twofold:  on the one hand  it  significantly reduces the computing time  and 
 on the other hand it allows us  to avoid the ambiguities related to the implementation of  condition (14) of Article {\it I}, that is 
%\begin{equation}\label{maxrep}
${\boldsymbol  \xi}_{max} \, \cdot  {\boldsymbol   \nabla }^2 \lambda_{max}\,  \cdot \, {\boldsymbol   \xi}_{max}  <0$
%\end{equation}
 , on a discrete relatively sparse grid. Therefore, since we need a point from which to start the numerical integration of LCS, we take the 
 {largest local maxima of  FTLE field as starting points.}
In principle,  we  should solve  for the curve defined by the condition ${\boldsymbol e}_0=  {\boldsymbol \xi}_{min} $ (Eq.(12) of Article {\it I} ) %(\ref{condizione2})
where we recall ${\boldsymbol e}_0$ is the tangent vector to the material line  and ${\boldsymbol \xi}_{min}$ the eigenvector of the Cauchy-Green  strain tensor corresponding to the smaller eigenvalue)  starting from each local maximum. However in chaotic systems the FTLE field exhibits a huge number  of local maxima and, in addition, the numerical evaluation of the matrix $\nabla {\boldsymbol \phi}_{z_0}^z(x_0,y_0)$ produces  a very discontinuous FTLE field. By integrating the above condition % Eq.(\ref{condizione2}) 
for each local maximum we  would find so many  structures that they  would confuse the physical information which  we want to extract. For these reasons it is necessary to define a criterion to adopt in order  to filter out the maxima that we consider not to be  physically significant.  It is clear that the larger the area around a local maximum, the more significant the maximum can be assumed to be. This criterion corresponds to take only those points that are absolute maxima of FTLE field within a  predefined area. Therefore, we seek   maxima of FTLE field and then, if the distance between two maxima is smaller than the  predefined maximization distance, we disregard  the maximum  with  the lower value of the  FTLE field. In other words the number of LCS  that we find
depends on the value that we choose  for the  maximization distance in the code. \\
{In order to clarify this criterion let us suppose that two large maxima of the FTLE field are very close to each other, i.e.  that their distance is smaller than the chosen maximization distance. In this case only  the largest  maximum is used as a starting integration point. In general this does not lead to a loss of  physical information since, if   two maxima are strong and are very close to each other, usually the LCS goes through both maxima  and thus the distinction between them is no longer necessary.
However, it is  also possible that two close maxima may give rise to different LCS: in this situation we miss one LCS because we keep only one maximum. In the following we  will illustrate in a specific case how the maximization distance can affect the resulting LCS and the physical information that we can obtain on the system.}

In our simulations we performed a series of tests to tune the value of the maximization distance  in order to find the optimal value that allows us to characterize the behavior of the system avoiding to have to deal with  too many structures. We used a resolution of $600$ points {in the $x$-direction and the number of points in $y$-direction is set  such as to have the same spatial resolution, i.e. $\Delta x = \Delta y$. \\  Another  critical parameter  is the {interval $z\, - \, z_0$ chosen in the computation of the Cauchy-Green} tensor. In fact, if this interval  is too small, we risk selecting structures that last for a too short $z$-interval:}  for example, we could find LCS also in a non chaotic region, since two KAM tori with different velocities could be seen as divergent trajectories if the evolution time of the system is too small.
  On the other hand, if the integration interval  is too long, the computational time grows and {it  may became very difficult to follow the eigenvectors of  the Cauchy Green tensor.} {Moreover, as this interval increases, LCS tend to converge to the corresponding invariant manifolds which, as stated in Article {\it I}, are characterized by a very convoluted structure. 
Choosing carefully this parameter we are able to find structures with relatively simpler patterns which describe the coherent behavior of the system on a shorter interval.} Finally, if the $z$ interval is too long, the LCS technique itself could be wrong because the LCS are computed using linear techniques  (see Eq.(5) of Article {\it I}) . %\ref{3=})).
In order to avoid the problems related to a possible  bad choice of the $z$ interval, the Cauchy-Green tensor has been computed by taking, in the periodic case, the numerical value of $\nabla {\boldsymbol \phi}_{z_0}^z(x_0,y_0)$
  every {$8 L_z = 4 \cdot 32 \pi$.  This means that the FTLE field is calculated  after every 4 $z$-loops and that we can  take, among the points that are maxima of FTLE field at the end of $z$-interval, those points that have after  each 4 $z$-loops a value of the  FTLE 
  larger than  the FTLE  mean value (computed with  the values taken at the grid points).  Since in our simulation the interval $z - z_0$ is 16 loops,  this corresponds to  three checks. With this check we ensure  that we  take as a starting point a point that repels particles at each time instant. In other words, we want to take a point that has a good repulsion property during the 16 loops of  the  simulation  interval, although this does not necessarily imply that the repulsion properties can be extended to longer integration intervals.

Finally, special attention has been  paid to the problem of noise arising  from the use of  a finite grid. This problem is enhanced in a chaotic system and is  mitigated here, {as mentioned before}, by  avoiding spurious maxima present only at the end of the $z-z_o$ interval and by filtering the LCS {by means of the criterion described above}.

\section{The $z$-periodic case} \label{resmagnper}

In this section we show the simulation results that we have obtained considering the Hamiltonian for the magnetic field line  trajectories at a fixed {physical  time}. {We choose} $t=415${, i.e.} before the onset of fully developed chaos.

\subsection{ Poincar\'e map}
As stated above, once the  {physical  time} has been fixed we can exploit the periodicity of the system along the $z$-direction and apply the  Poincar\'e map technique by {plotting the magnetic line  intersection points  in the $x$-$y$ plane after each periodicity interval in $z$ starting from a given initial value of $z$ that defines the section chosen}. This kind of plot is very useful since it provides information on the topological aspect of the magnetic field configuration, identifying the regions where the {trajectories  are regular  and those where  they are chaotic}. Although the  Poincar\'e map is fundamental to study a time periodic dynamical system, the LCS technique allows a more refined analysis, as it makes it  possible {to  further partition the regions characterized by a chaotic behavior into sub-regions where trajectories have a qualitatively different behavior on the  time intervals  which characterize the LCS. }  
Here we focus on two sections,  at $z=0$, and $z=L_z/2$  respectively, and restrict the integration domain in the $x$-direction to the region in between the two initial resonant surfaces where chaos develops first i.e., to  $0<x<0.8$ and $-2\pi<y<2\pi$. The corresponding plot for $z=0$ is shown in Fig.\ref{PP1}. We note that a chaotic region exists  between the {two island chains} that correspond to the initially imposed perturbations. However regular regions survive in the chaotic sea: in 
 particular  the regular region having {$ m = 3  $} indicated in the figure splits the domain in two sub-domains.

\begin{figure}[htb]
\centering
\includegraphics[height=8cm,width=9cm]{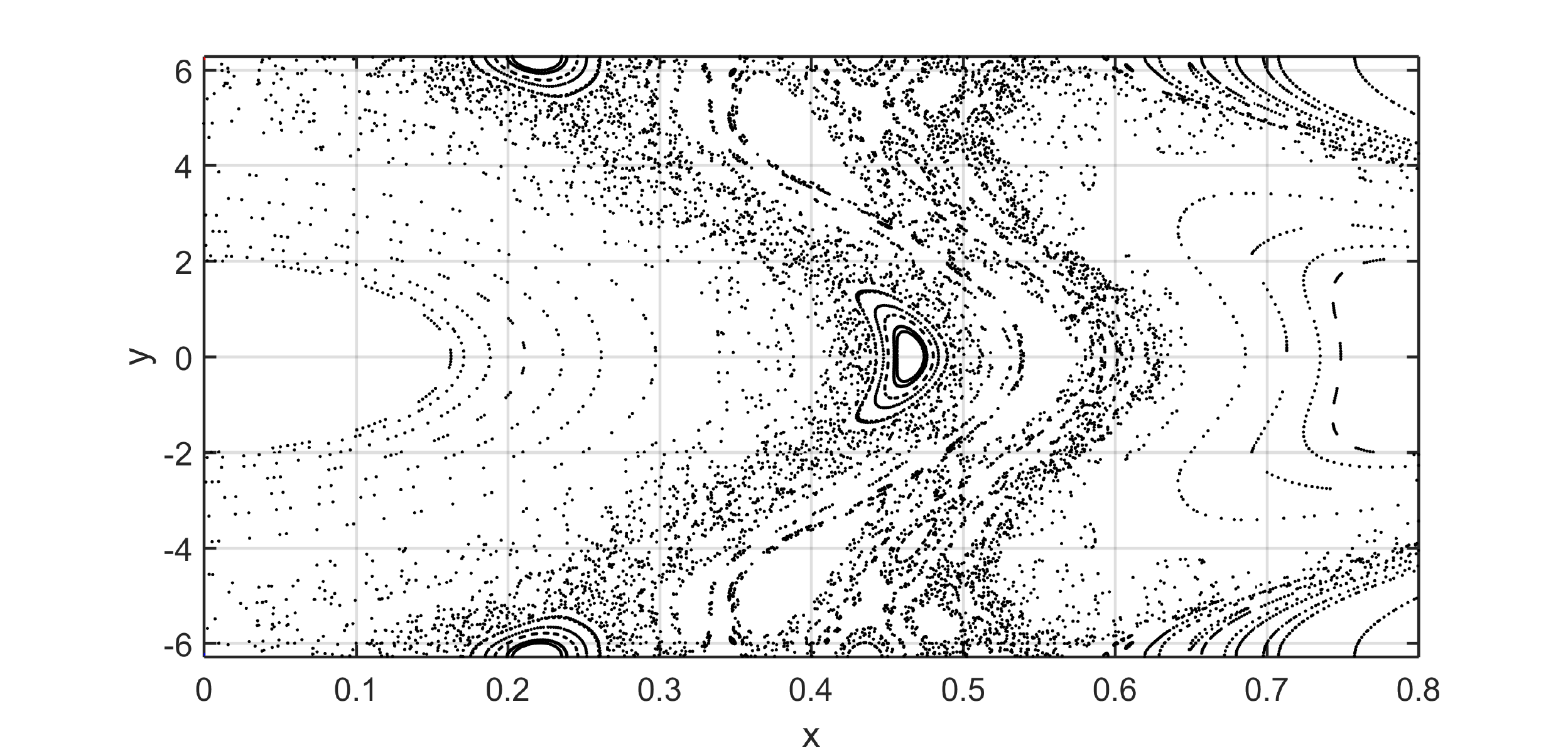}
\caption{ Poincar\'e map at $z=0$ of the magnetic configuration taken at $t=415$. The map has been obtained evolving $150$ initial conditions for $100L_z$. The initial conditions are uniformly distributed in the $x$-direction in the interval $[0,0.8]$ and have $y = - 2 \pi$. The white regions %in the map 
correspond to trajectories  (i.e. to magnetic field lines)  on regular surfaces in   the extended $x,\, y, \, z$ phase space. The larger  ones at the edge of the domain are the magnetic islands corresponding to the initial  perturbations, while the regular region, that for  $y=-2\pi$ is located approximately between $x=0.35$ and $x=0.4$, 
corresponds to a chain having {$ m = 3  $} and splits the chaotic region in two sub-domains.}
\label{PP1}
\end{figure}

\subsection{Lagrangian coherent structures}

To find the LCS  it is first necessary to compute the field of the finite time Lyapunov exponents (see Secs. V and V B of Article {\it I})  and then
proceed  with the trajectory integration starting from the points with the largest eigenvalue $\lambda_{max}$.
The FTLE field is shown in Fig. \ref{FTLE}.
\begin{figure}[htb]
\centering
\includegraphics[height=8cm,width=9cm]{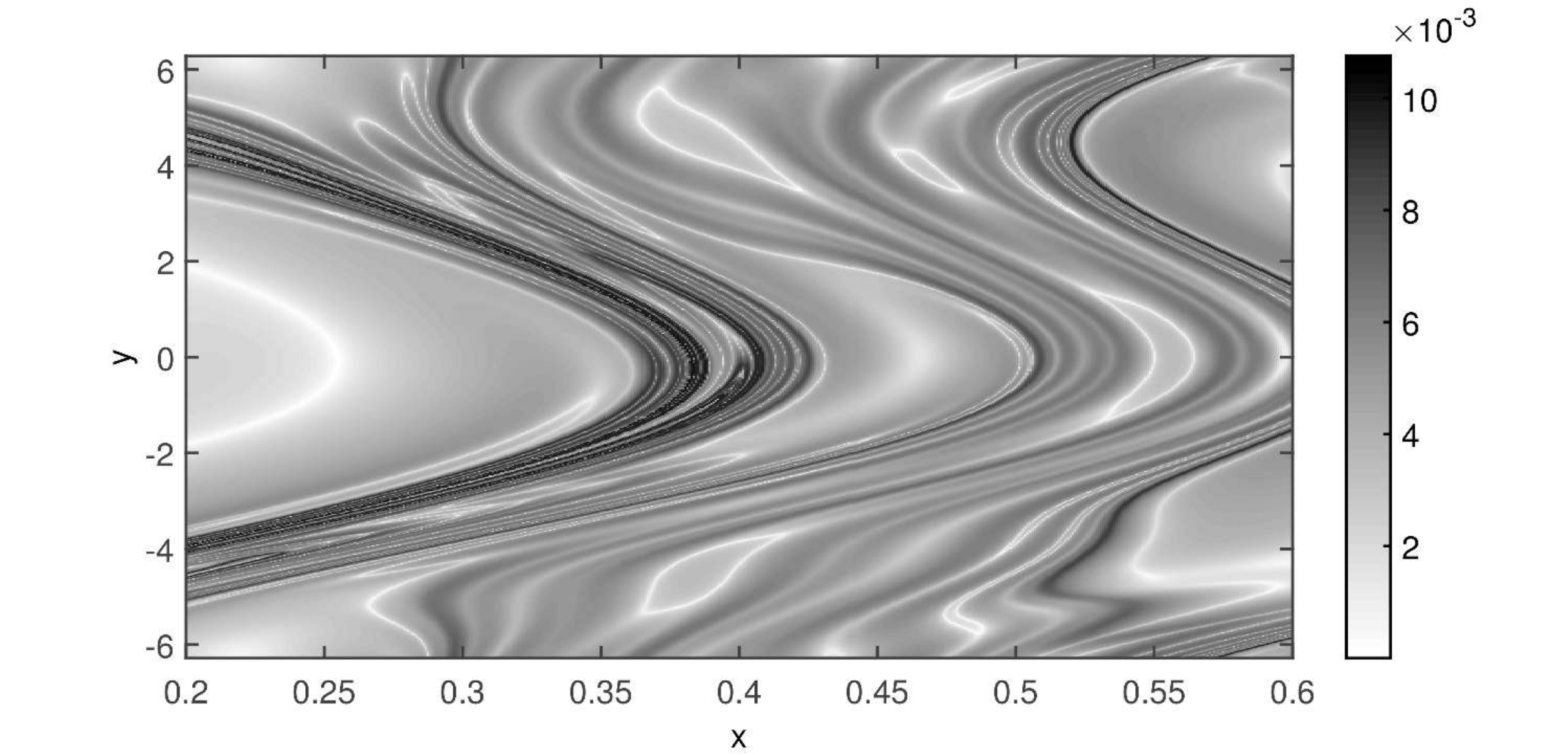}
\caption{FTLE field  for the Hamiltonian at $t=415$,  on the plane $z=0$.  $500$ points in $x$, and $7850$ in $y$ have been used  so as to have the same resolution in both the directions.
The darker shading   corresponds  to larger values of the eigenvalue  $\lambda_{max}$ of the Cauchy-Green tensor. }
\label{FTLE}
\end{figure}
{In the following we restrict our search  to repelling LCS since  we can exploit the space-time reflection symmetry $y \rightarrow - y$, $z \rightarrow - z$
introduced in Sec. VIA  of Article {\it I} in order  to find the attractive LCS  which   are thus obtained  by mirror  reflection of  the repelling ones  with respect to the $y =0$ axis.}\\          
{ In Fig. \ref{PPLCS} the LCS  that we have  identified  with the numerical procedure described} in Sec.\ref{intmet}
{are overplotted on the  Poincar\'e map. The repelling (attractive) structures are drawn in red (blue).
We recall that in the small amplitude linear phase the two  perturbations with different helicities
evolve independently from each other and each of them  induces
a magnetic island chain around its resonant surface. {If the Hamiltonian does not depend on $z$,} these islands  are delimited by the separatrices that are  formed through the smooth connection of stable and unstable manifolds and that act  as  barriers}. {Here the smooth connection between stable and unstable manifolds is broken  since the magnetic configuration does not correspond to an autonomous dynamical system}. The footprint of the breaking  can be recognized in Fig.\ref{PPLCS}    close to the regular regions corresponding to the initial perturbations   in the lobe-like shape  that the LCS exhibit  when approaching the edge of the domain. Since LCS mark the most repelling (attractive) material lines, they tend to follow the stable (unstable) manifolds that in non the autonomous case  continue  to intersect the unstable (stable) manifolds.  
In  Fig.\ref{PPLCS} we mark with green arrows the most visible intersections that give rise to the lobes. In principle, the intersection should continue to generate a complex tangle but our numerical integration can not follow the manifold  oscillations indefinitely.
\begin{figure}[htb]
\centering
\includegraphics[height=8cm,width=9cm]{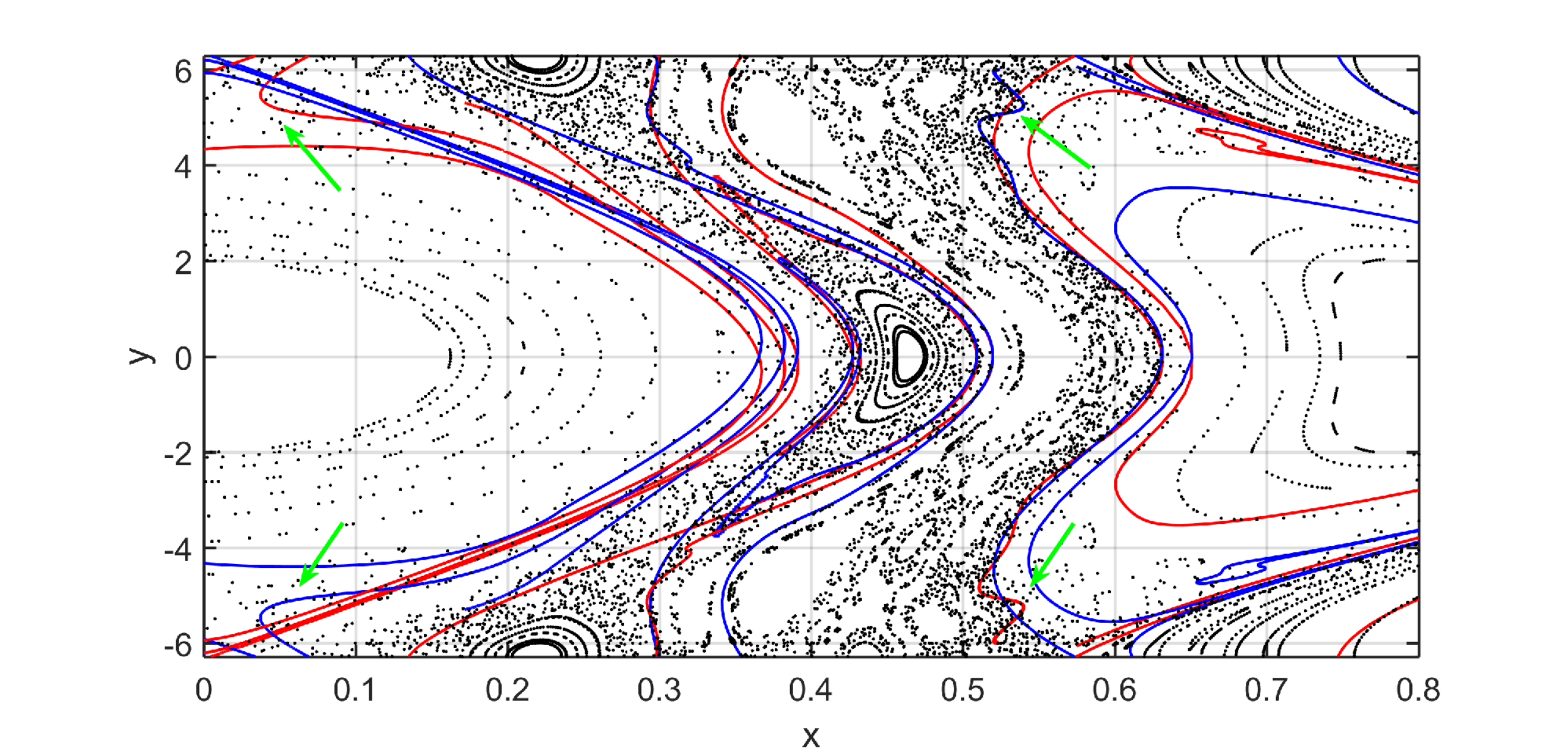}
\caption{Most important LCS overplotted on the  Poincar\'e map at $z=0$ and $t=415$.
 The repelling (attractive) structures are drawn in red (blue). The green arrows indicate where the lobes and the tangle form.}
\label{PPLCS}
\end{figure}
\noindent

In order to test the robustness of these LCS as barriers we performed a series of trajectory integration of magnetic field lines  considering an initial {set} of $20$ initial conditions at a given   position and letting these trajectories evolve for $80 L_z$. 
All the initial conditions (i.c.) are localized into a radius of 0.003. Then we plot their position
in the $x$-$y$ phase space at  every crossing of the  the $z=0$ section on which we have calculated the LCS. Figs. \ref{barriera3_4}-\ref{barriera21_24} confirm that the LCS that we have found act as strong barriers, since 
 there is no flux through them on the considered time-span  unless we consider regions with lobes and tangles. The location of the initial conditions is {marked}  by an arrow in the figures.  In the left panel of Fig.\ref{barriera3_4}, by taking the initial
 conditions very close to the KAM surfaces that are still present in the chaotic sea that forms between the two main magnetic islands, we see how the LCS confine the evolution of these trajectories. On the contrary, in the right panel of Fig.\ref{barriera3_4}  we set the initial conditions very close to {a}
 repelling LCS. In that region lobes and tangles are expected, although they are not visible due to the small resolution: with  the adopted resolution we are able to follow the manifolds of the main islands, corresponding to  the  {$m = 1$}  mode but we cannot follow the manifolds of the smaller islands. {In this region, according to the lobe dynamics }briefly recalled  in Sec. III  of Article {\it I}, particles can cross the barriers.
In the remaining Figs. \ref{barriera13_15},\ref{barriera21_24} the role of different LCS is  again tested  using  the same technique.

\begin{figure*}[htb]
\centering
\includegraphics[height=7cm,width=7cm]{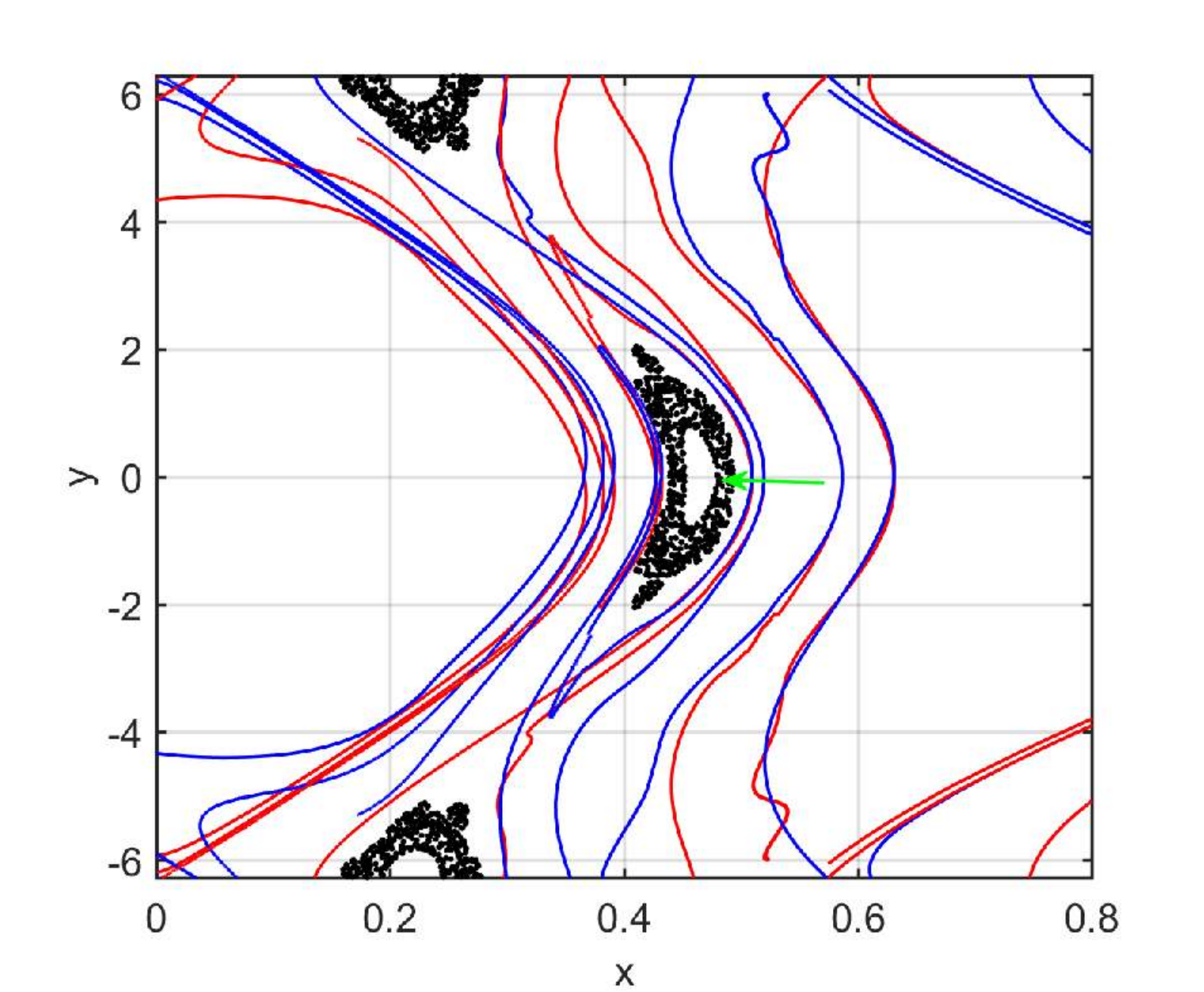}
\includegraphics[height=7cm,width=7cm]{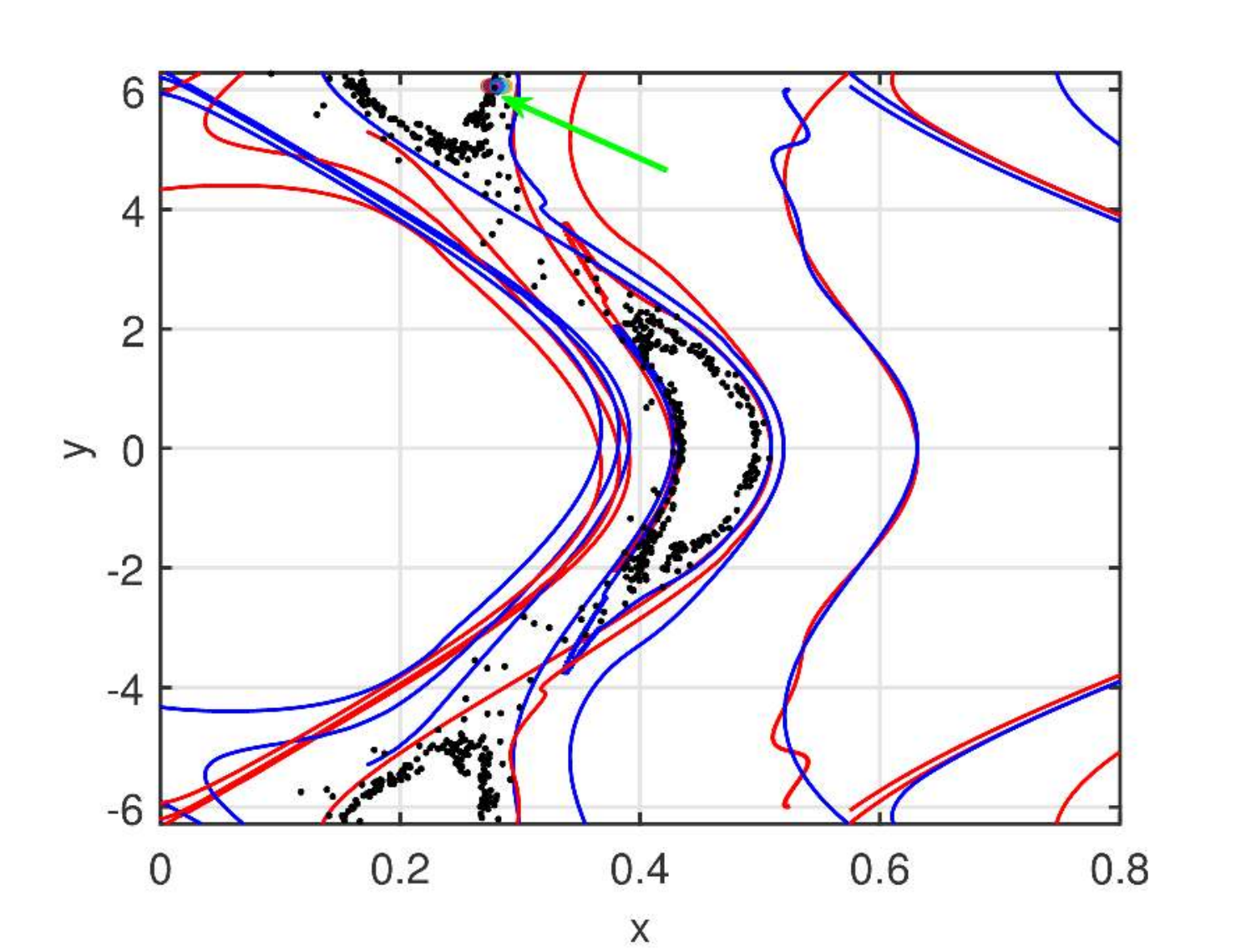}
\caption{LCS obtained using the Hamiltonian at $t = 415$.  The  left panel  shows  that the initial conditions taken in the regular region, bounded by hyperbolic LCS,  remain confined  inside this region. In the right  panel  the initial conditions are  very close to a repelling LCS and, therefore, some particles  escape according to the lobes dynamics. The location of the initial conditions is marked by an arrow.}
\label{barriera3_4}
\end{figure*}

\begin{figure*}[htb] 
\centering
\includegraphics[height=7cm,width=7cm]{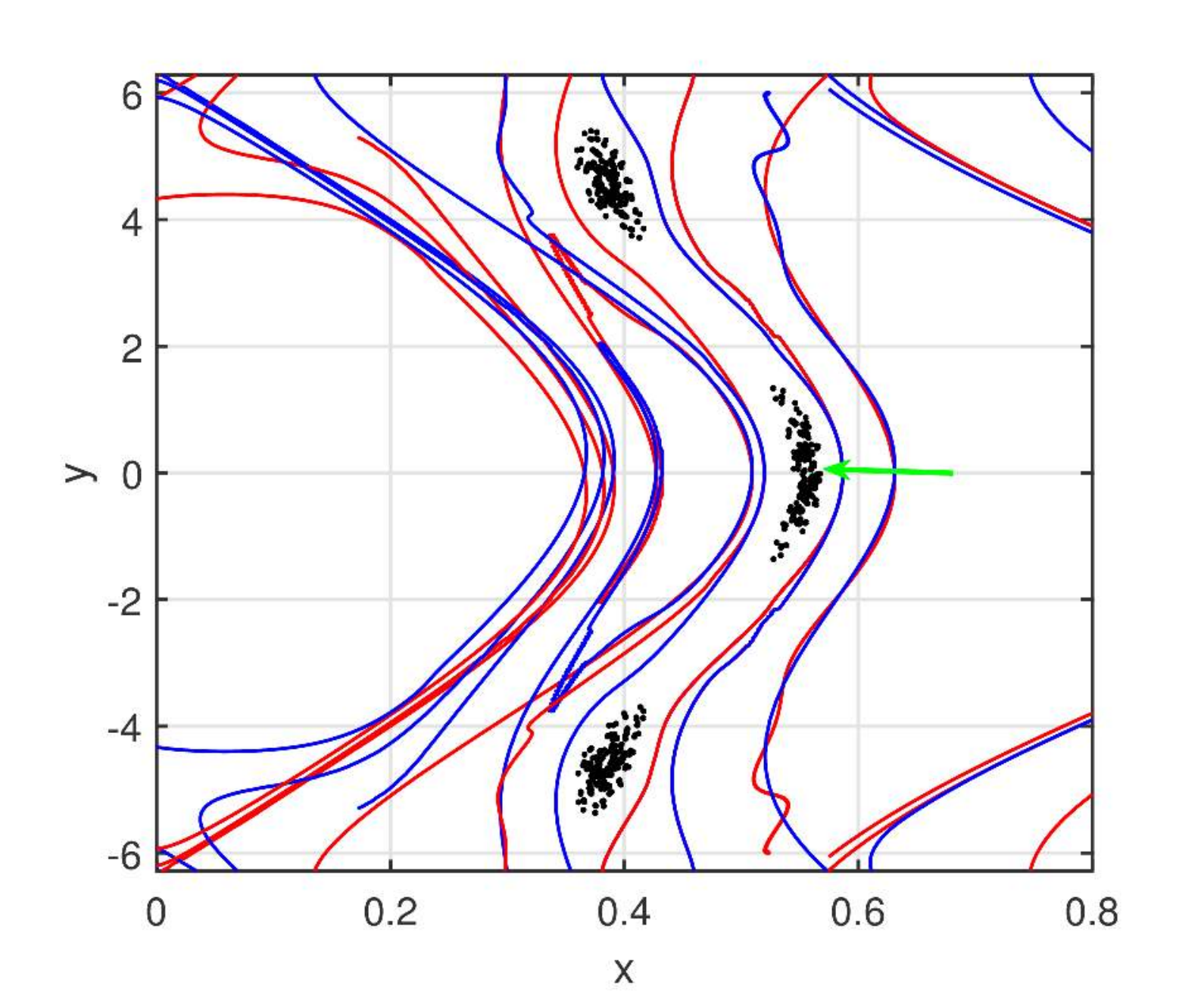}
\includegraphics[height=7cm,width=7cm]{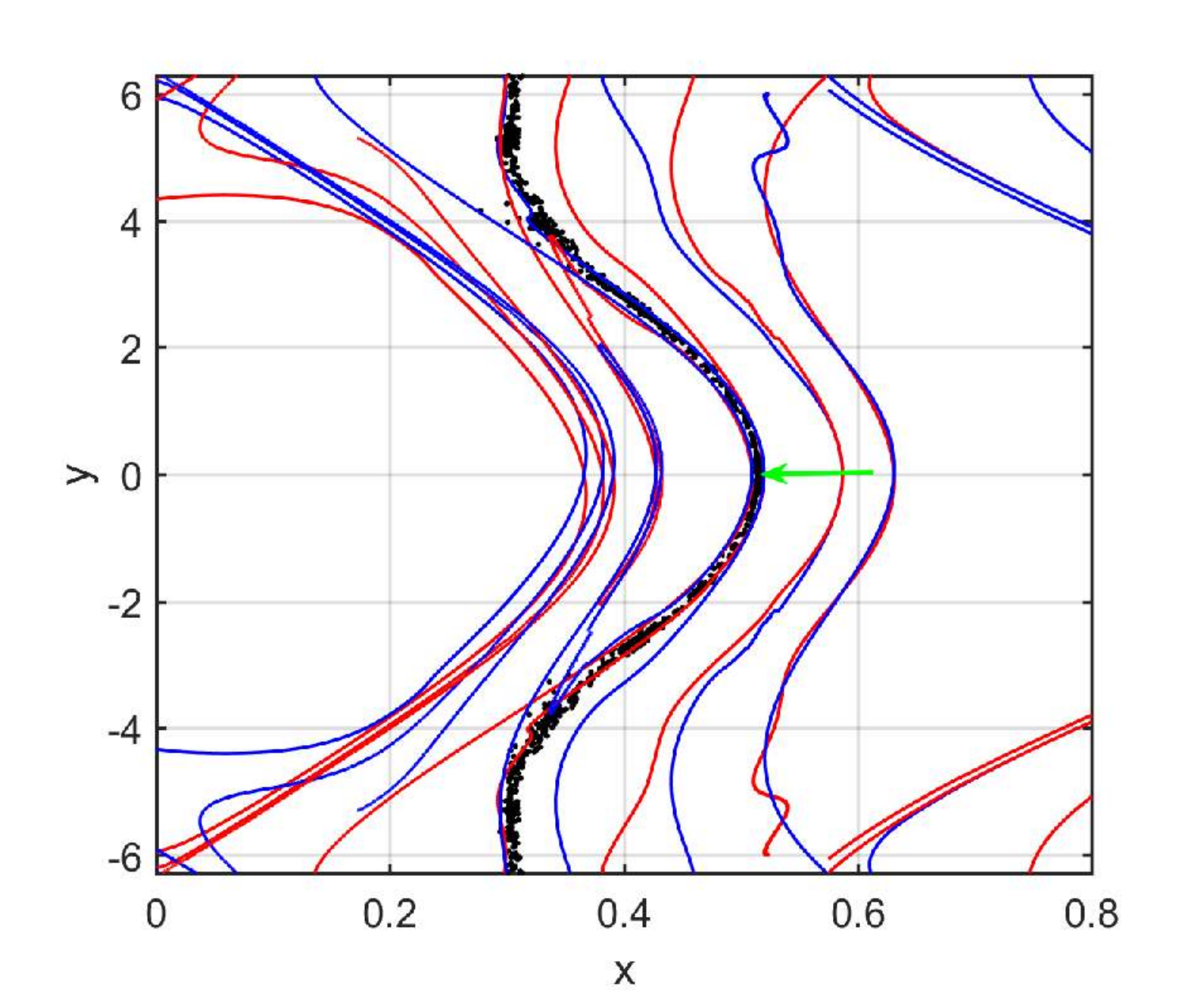}
\caption{LCS obtained for the Hamiltonian at $t = 415$. Both figures show how the drawn LCS  act as barriers. Note that the marked  trajectories  belong to two different regions:  in order to make the visualization easier  in the right frame we have  deleted the LCS that confine the set of initial conditions in the  left frame. The location of the initial conditions is marked by an arrow.}
\label{barriera13_15}
\end{figure*}

\begin{figure*}[htb]
\centering
\includegraphics[height=7cm,width=7cm]{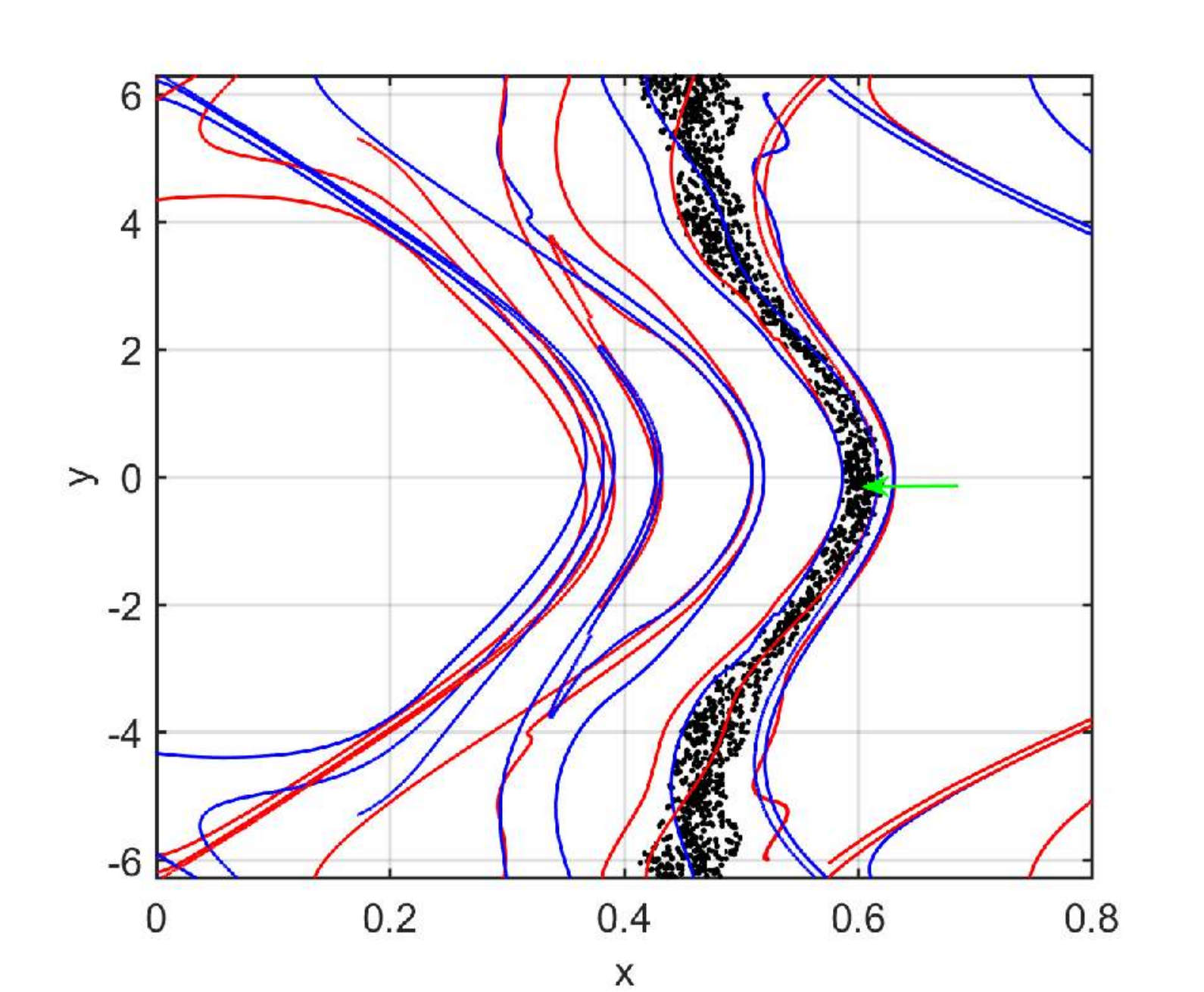} 
\includegraphics[height=7cm,width=7cm]{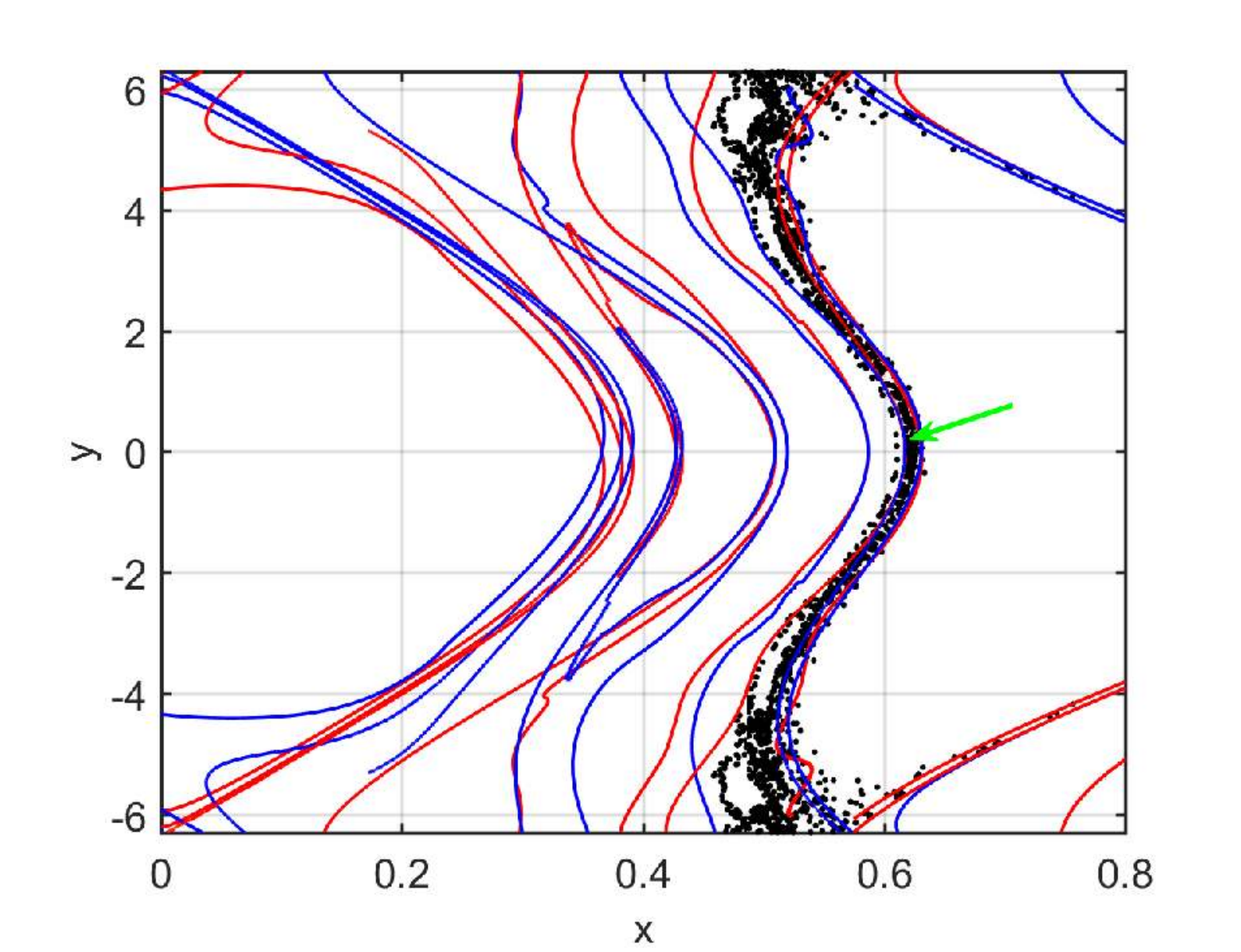}
\caption{LCS obtained using the Hamiltonian at $t=415$  Both figures show how the drawn LCS  act as barriers.  Note that the  figure in the right frame has been obtained reducing the value of the  maximization distance  with respect to  that in the left  frame. 
A new  LCS  arises and it splits the chaotic domain in the left  frame  in two sub-domains. Both  domains are  chaotic but they cannot communicate. This underlines the fact that if we  take  a smaller value of the  maximization distance  we can find additional transport barriers. The location of the initial conditions is marked by an arrow.}
\label{barriera21_24}
\end{figure*}

Examining  the plots of the LCS  shown in the figures we note that   the repelling LCS, red lines in  Fig.\ref{PPLCS} (and similarly  the attractive LCS, blue lines)  appear  not to be periodic  in  the $y$ direction.  Actually this is a numerical effect  related partly to the  size of the integration grid and partly to the setting of the maximization distance described in Sec.\ref{intmet}.
In fact}, decreasing the maximization distance used when selecting  the FTLE maxima,  additional LCS arise among which are those that match, at the edges of the $y$ domain, the structures shown in the  plots.

\section{The $z$-non periodic case} \label{res}

In this section we consider the case of a magnetic field that evolves  in   physical time.  This implies that a  charged particle moving  in the plasma 
sees a  time varying magnetic field  during its motion. The  main questions  that we intend to address  are: \,how do  the LCS change with  physical  time, 
how different are  the  new LCS  from those  found at the  fixed physical  time $t=415$ and finally   if and how particles  can cross LCS constructed for  particles with  a different velocity.

\begin{figure}[htb]
\centering
\includegraphics[height=8cm,width=9cm]{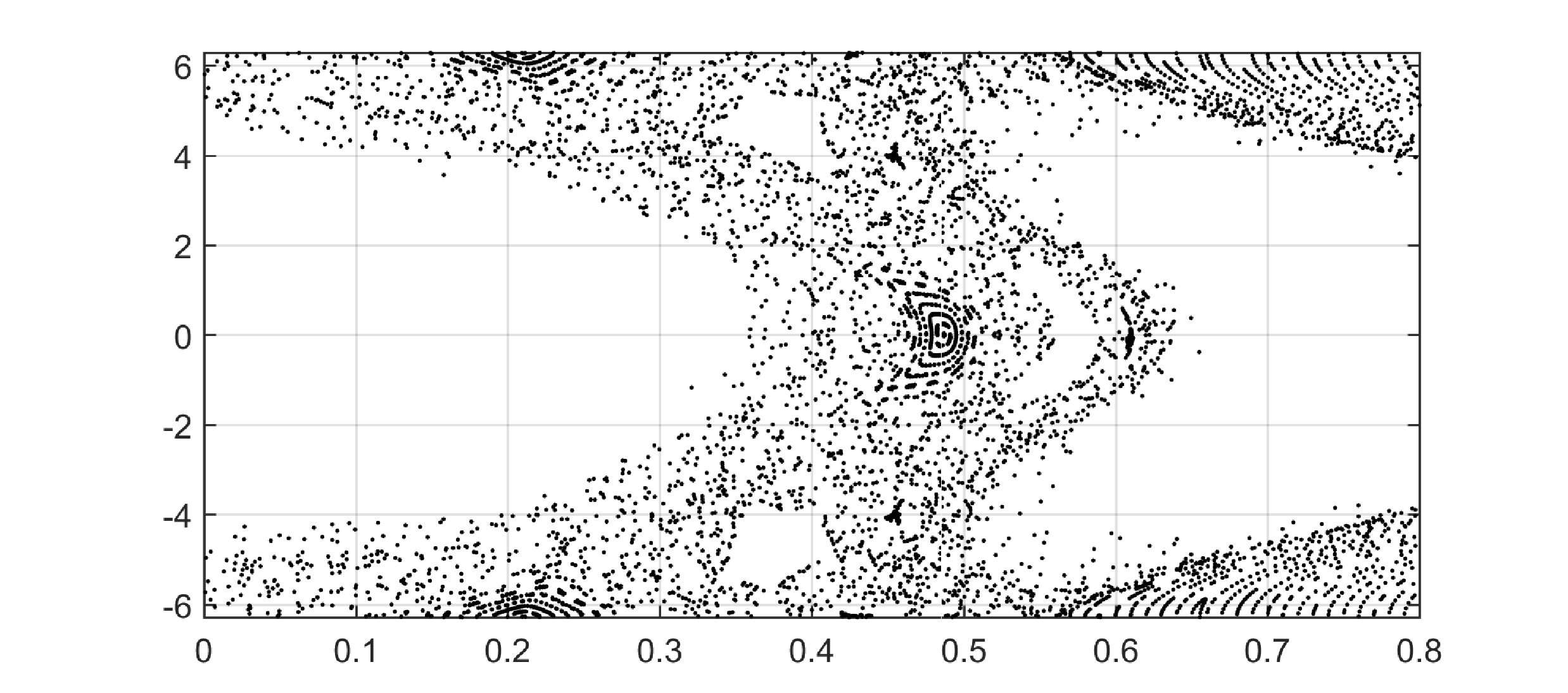}
\caption{Poincar\'e map at $z=0$ and $t=425$.}
\label{PP2}
\end{figure}

We examined a time interval extending over $10$ normalized units  from $t_1 = 415$ to $t_2 = 425$. Chaos has developed during this time interval as  shown by the Poincar\'e map   in Fig.\ref{PP2} for  $t=425$ i.e. at the end of the interval. 

We adopt the simplified model, where particles move   with a constant   velocity ${V}$ along the $z$-direction  only, described in Sec. VIB  of Article {\it I}. 
Then the new   Hamiltonian  is given by the modified   flux function  
\begin{equation}\label{psiV}
\psi_{\cal V}(x,y,z)\,\, \equiv \,\,  \psi(x,y,z, t = (z-z_o)/{ V}). \end{equation}
 (see Eq.(21) of Article {\it I})  where the physical time dependence is chosen according to Eq.(\ref{interplaw})  and the  dependence of  $\psi_{\cal V}$   on the  $z$ variable combines  the spatial  and the time dependence of the magnetic configuration,  as seen by a particle streaming with velocity $V$ along a field line.
We use this Hamiltonian to calculate LCS for different values of the velocity $V$.  Note that, although simulations with different values of $V$ have been performed, unless specified, the LCS  shown in  the figures  are those for  particles with velocity 
$ V=1000$. With this value the particles perform $10 \, z$-loops in one time interval.
  This is a compromise between having a magnetic field that does not evolve too fast during the motion of particles and 
being able to show the  dependence  of the LCS on the velocity $V$ and  to investigate whether  or not the LCS computed for a given velocity $V$  act as barrier also for particles with   different velocities. In the following we will focus on particles with a positive velocity.  {We stress  that in this model the $z$ periodic case corresponds to the assumption that  the particles move  with infinite speed and  thus experience a fixed magnetic configuration.}

First we note that  if we keep  the number of  $z$-loops fixed   the LCS that are   found   with  increasing  velocity  turn out to be  similar to those found in the periodic case, as   expected  when  the particles travel time is much shorter  than the time over which the magnetic field changes.

Thus, in order to show in a more evident way the LCS  in a time evolving magnetic configuration as seen by a particle with velocity $V$,  we integrate the Hamilton equations (\ref{psin1})  with the flux function $\psi_{\cal  V}$ in Eq.(\ref{psiV})  over  fixed time intervals i.e.,  in terms of  the variable $t $  instead of $z$ using the relationship introduced  below Eq. (21) of Article {\it I}, that is  $t=(z-z_0))/V$. The change of variable from $z$ to $t$ and the definition of the time intervals need to be performed differently when computing repelling and when computing attractive LCS.

\subsection{Repelling LCS}

{For the calculation of} repelling LCS we relate  $ t $  to $z$ in the Hamilton equations such that $t- t_0 = (z - z_o) /V$

{First we show how LCS evolve in time i.e. we calculate LCS at time $ t = {\bar t}_0$ and position $z_0$, then we "follow" the structures computing them at time ${\bar t}_1$ and position $z_1$, at time ${\bar t}_2$ and position $z_2$, and so on}. To do this we set ${V}=1000$ and  choose the initial particle position, i.e. $z_0 = 0$, for all particles. Due to the long computational time and to the fact that the adopted method uses linear techniques, we choose to evaluate the LCS integrating the initial conditions for a maximum of $20 ~z$-loops. For a velocity ${V}=1000$, $20~z$-loops correspond approximately to $\Delta t =2$. In order to investigate how the time-dependent magnetic field can affect {the evolution of}  the LCS, we evaluate the structures at different   times ${\bar t}_n$,  starting  from ${\bar t}_0=415$  since the time-independent analysis has been carried out at $t_1= 415$. We  divide the interval $  [{\bar t}_0 = 415,  {\bar t}_0 + \Delta t = 417]$ into  sub-intervals of duration  $\delta t = 0.1$.  After each $\delta t$ we  calculate the LCS. Thus we obtain  a set of LCS at the times  ${\bar t}_n = 415, 415.1, 415.2$ up to $417$.  This allows us to determine how the LCS  computed at ${\bar t}_0 = 415$ {evolve in time}.

Using these LCS data, we show  how particles initially separated by a repelling LCS evolve in such a way  that they remain apart and do not cross the LCS itself as it evolves in time.

Particles with different velocities have different trajectories and, therefore, different LCS. We {investigate} the dependence of these structures with respect to $V$. As can be seen  from the simulation results, LCS act locally, i.e an exponential departure from a repelling LCS is not observed. Initial conditions  feel
the "repulsion" of a repelling LCS only when they are  very close to it. Due to this local influence, two sets of i.c. divided by a repelling LCS evolve initially  in such a way  as to maximize the distance from the repelling LCS, e.g. see Fig. (\ref {tempi_intermedi}). After this first stage they have different evolution.

{\subsection{Attractive LCS}}

{For the attractive  LCS we relate  $ t $  to $z$ in the Hamilton equations such that $t - t_{end} = (z - z_{end})/V$. As explained in Article {\it I} we compute the attractive LCS as repelling LCS of the backward time dynamics. We show how attractive structures affect particle dynamics and how essential they are  in order to understand the transport features of the system. Looking only at the repelling LCS we can 
   have only a partial understanding of the dynamics, e.g. we are able to say that two sets of i.c. divided from a repelling structure evolve in order to stay apart, but if we want to know also how "fast" are the mixing  phenomena for those i.c., we need to calculate the attracting LCS. Following these considerations, we think that attractive LCS give a more intuitive description of the dynamics. They offer an intuitive understanding about how a big set of i.c. evolves.
We remember that, when we evolve the system from  $ t_{end} = 417$ to $t_{end} - \Delta t = 415$ to compute the attracting LCS, the structures are those corresponding to $t_{end}=417$ and they describe the behavior of particles at the time $t = 415$.  Finally, we remark that also attractive LCS act as transport barriers. In  the numerical results section, we exploit this fact to show that particles with  velocity $V_1$ can cross barriers obtained considering a different velocity $V_2$. }

\bigskip

\subsection{Numerical results}

\begin{figure}[htb]
\centering
\includegraphics[height=7cm,width=9cm]{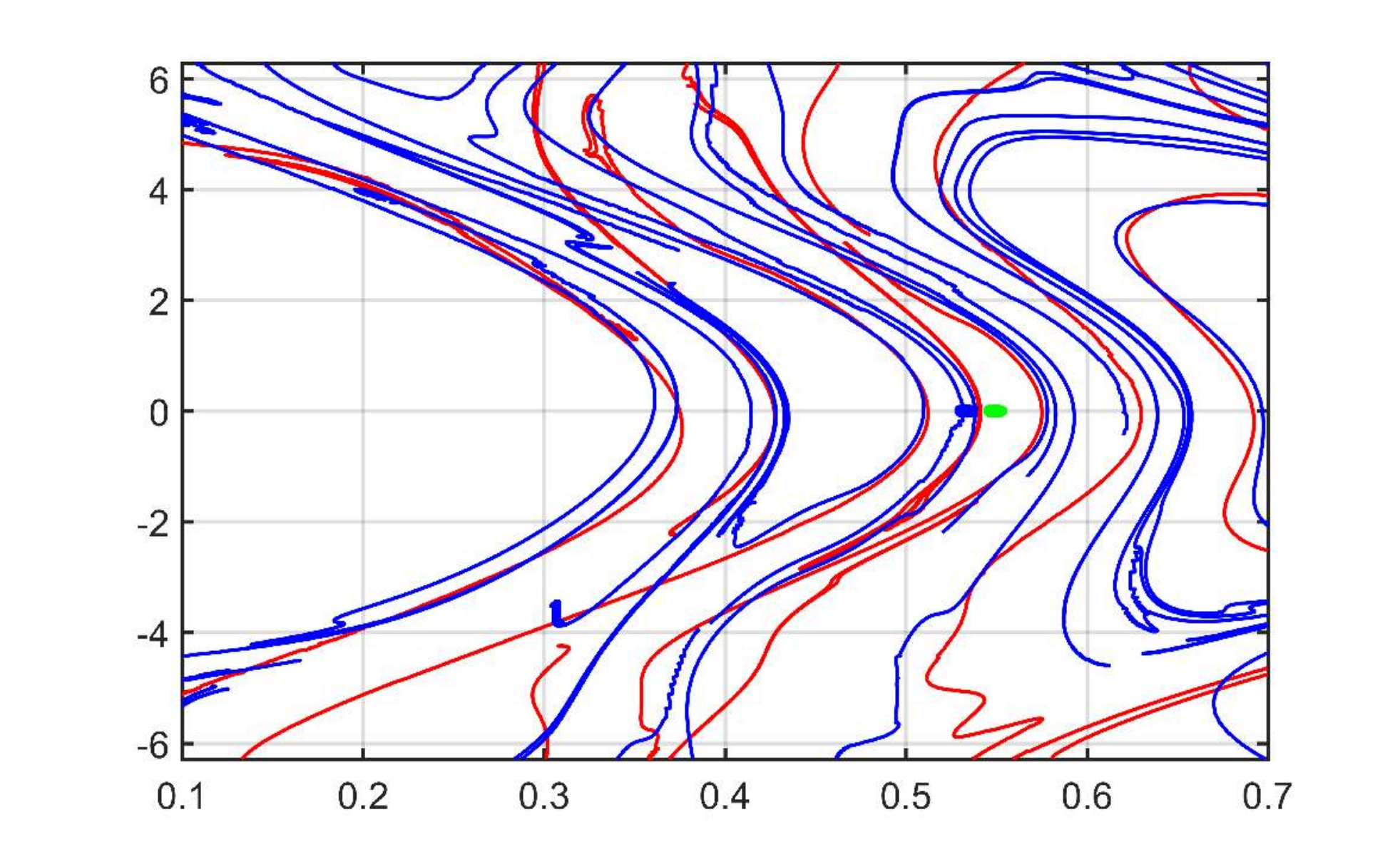}
\caption{Initial conditions split in two groups by a repelling LCS (red). 
Repelling LCS have been obtained with the system evolving from $t=415$ to $t=417$, and attracting LCS from $t=417$ to $t=415$. }
\label{tempo_0}
\end{figure}

\begin{figure*}[h!]
\centering
\includegraphics[height=8cm,width=7.5cm]{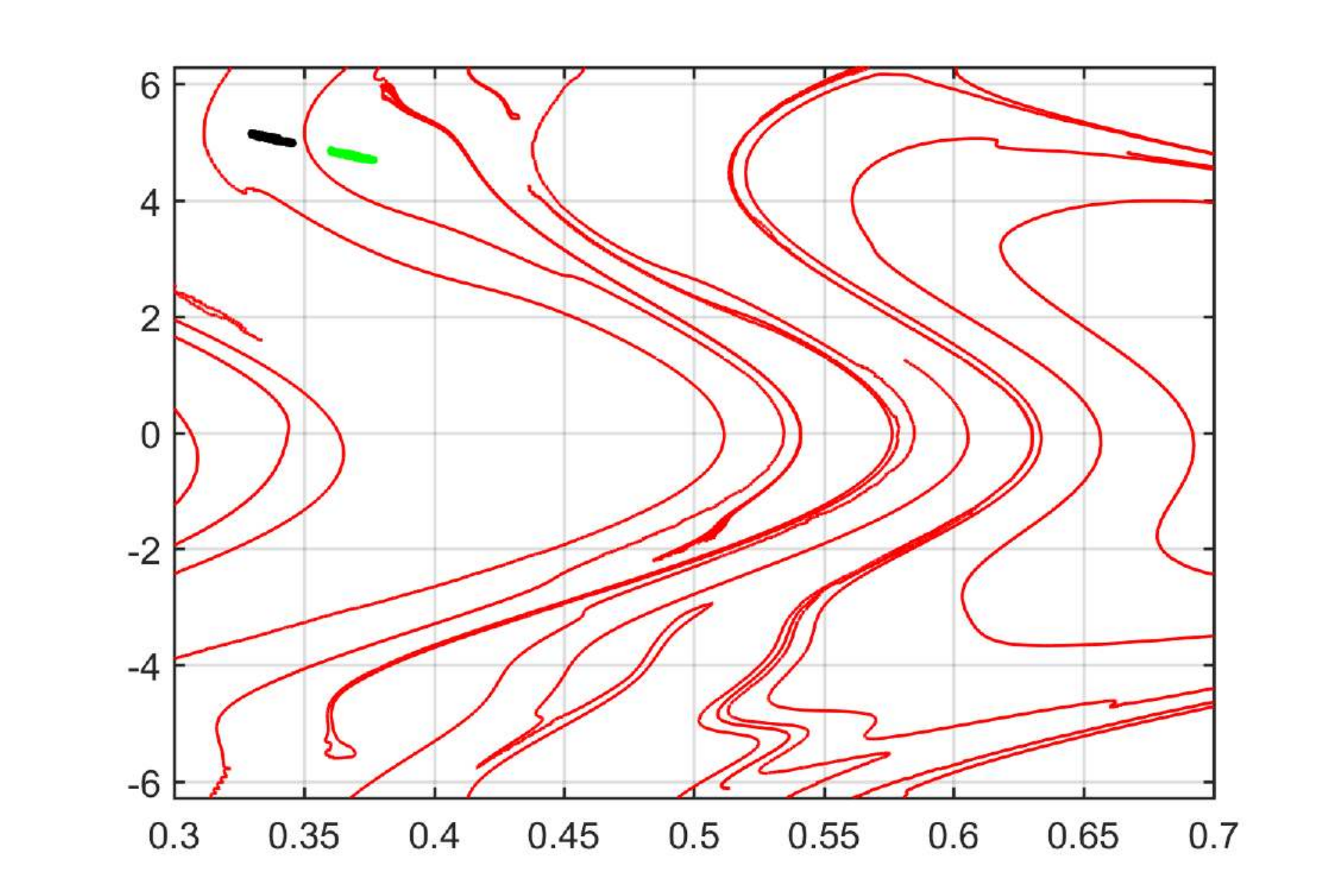}
\includegraphics[height=8cm,width=7.5cm]{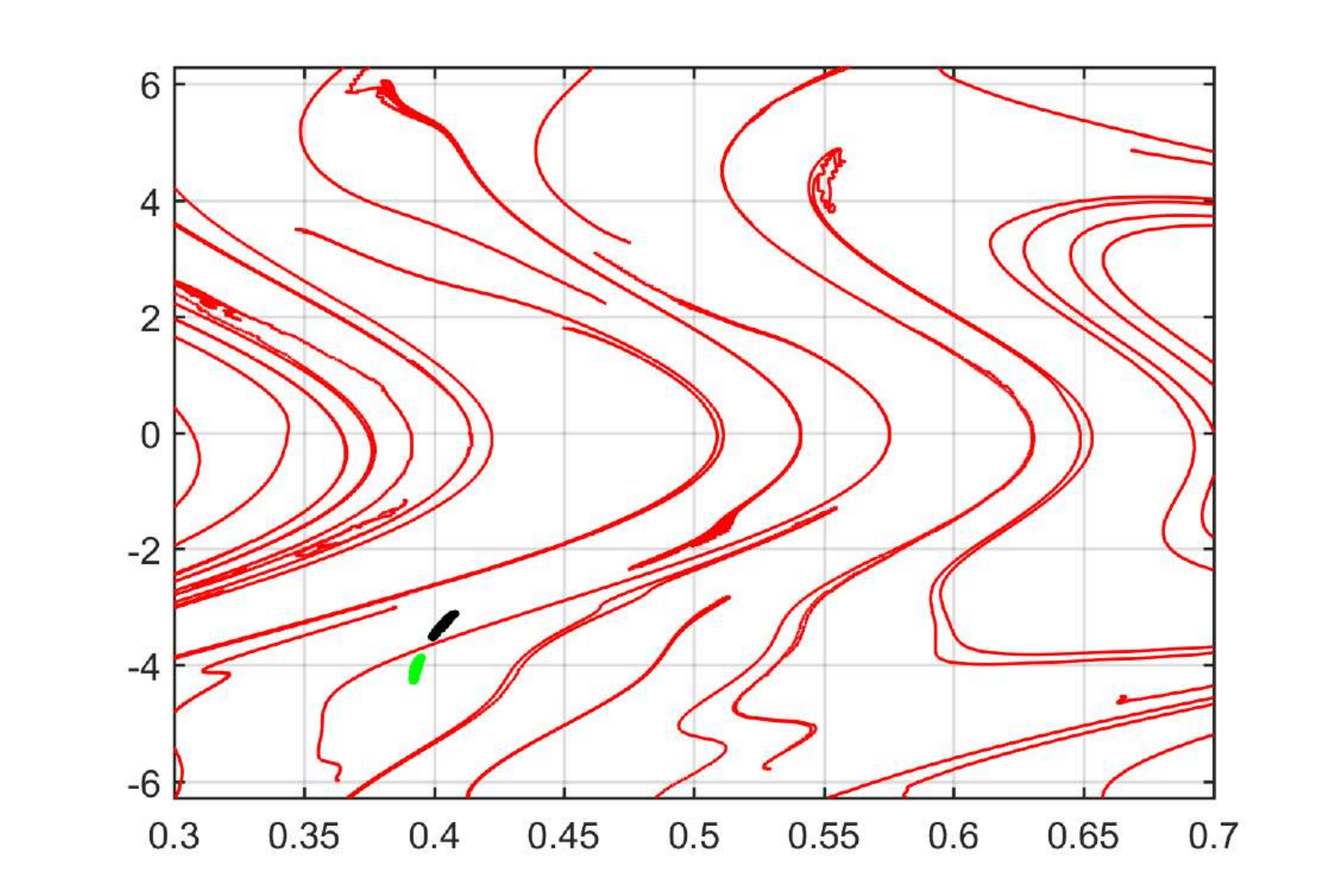}
\caption{Evolution of  the initial conditions and of  the LCS of Fig. \ref{tempo_0} at $t=415.1$ (left panel) and $t=415.2$ (right panel). The LCS have been obtained with the system that evolves from $t=415.1$ to $t=417.1$ (left panel) and  from $t=415.2$ to $t=417.2$ (right panel). }
\label{tempi_iniziali}
\end{figure*}

\begin{figure*}[h!]
\centering
\includegraphics[height=8cm,width=7.5cm]{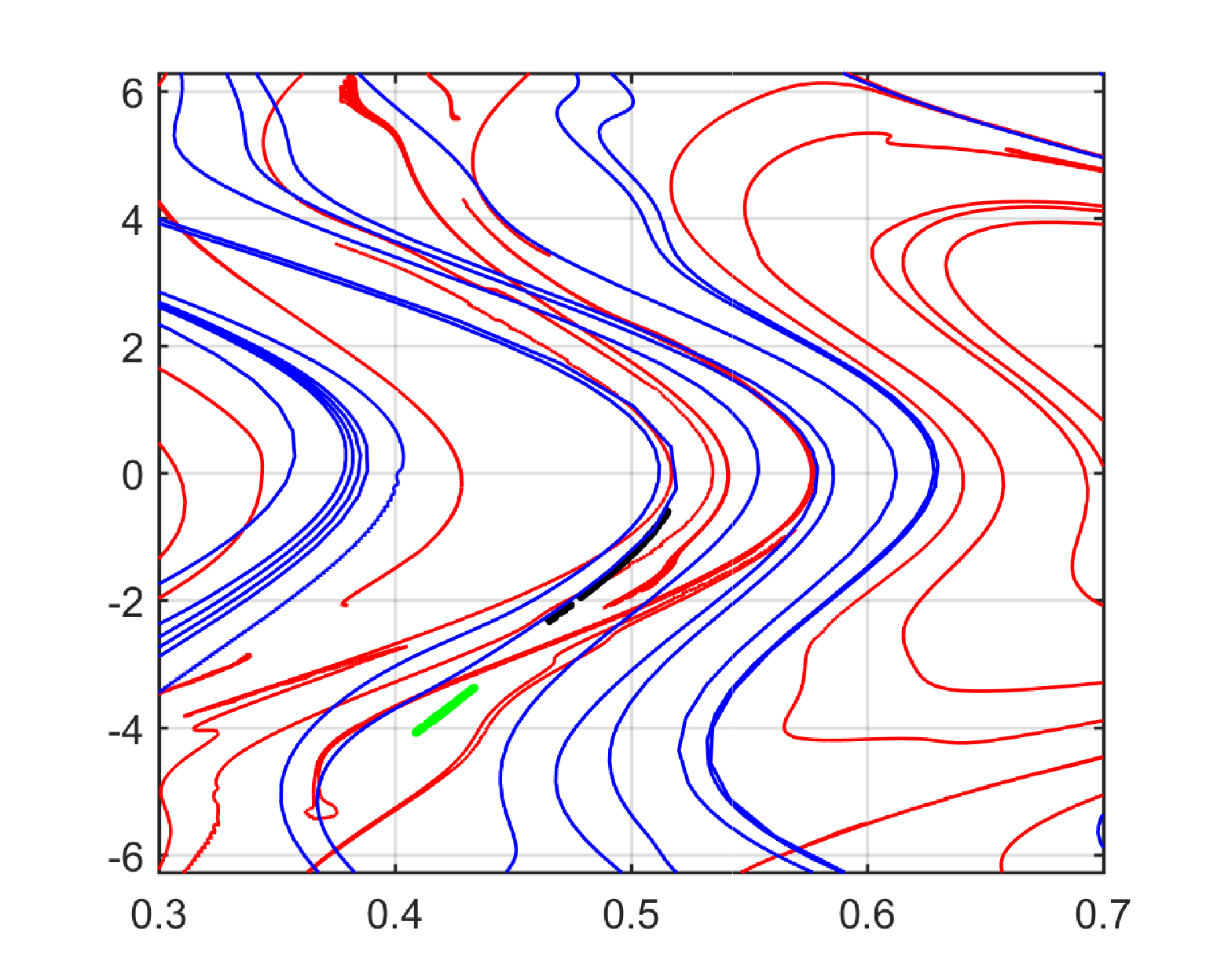}
\includegraphics[height=8cm,width=7.5cm]{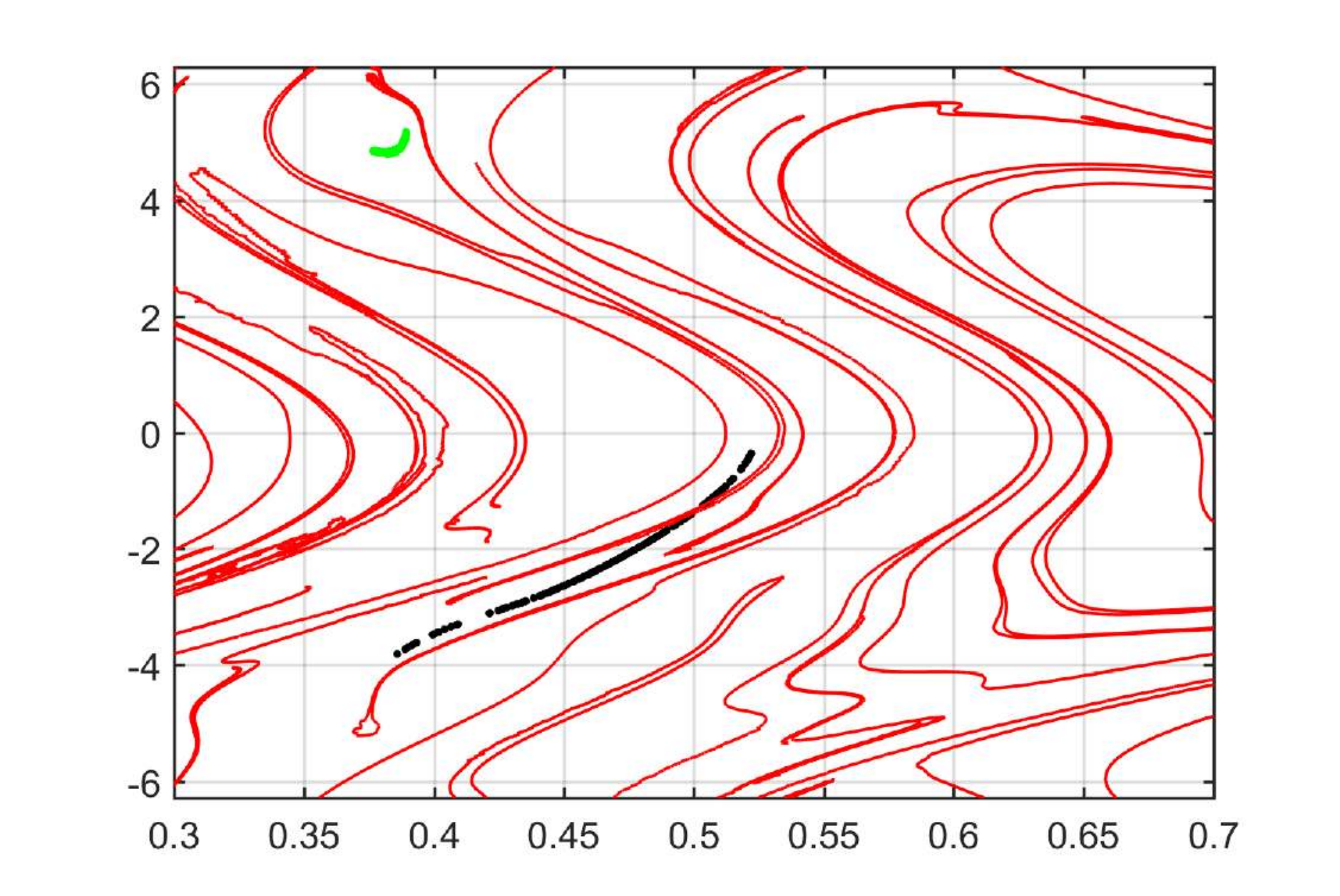}
\caption{Evolution of  the initial conditions and of  the repelling LCS (in red) of Fig. \ref{tempo_0} at  $t=415.5$ (left panel) and $t=416 $ (right panel). The repelling LCS have been obtained with the system that evolves from $t=415.5$ to $t=417.5$ (left panel) and  from $t=416$ to $t=418$ (right panel). In the left panel, the attracting LCS (blue curves) have been obtained with the system that evolves from  $t=415$ to $t=415.5$. This is just to show how particles arrange themselves along attracting LCS computed for the same time interval of the particles evolution.}
\label{tempi_intermedi}
\end{figure*}

\begin{figure*}[h!]
\centering
\includegraphics[height=8cm,width=7.5cm]{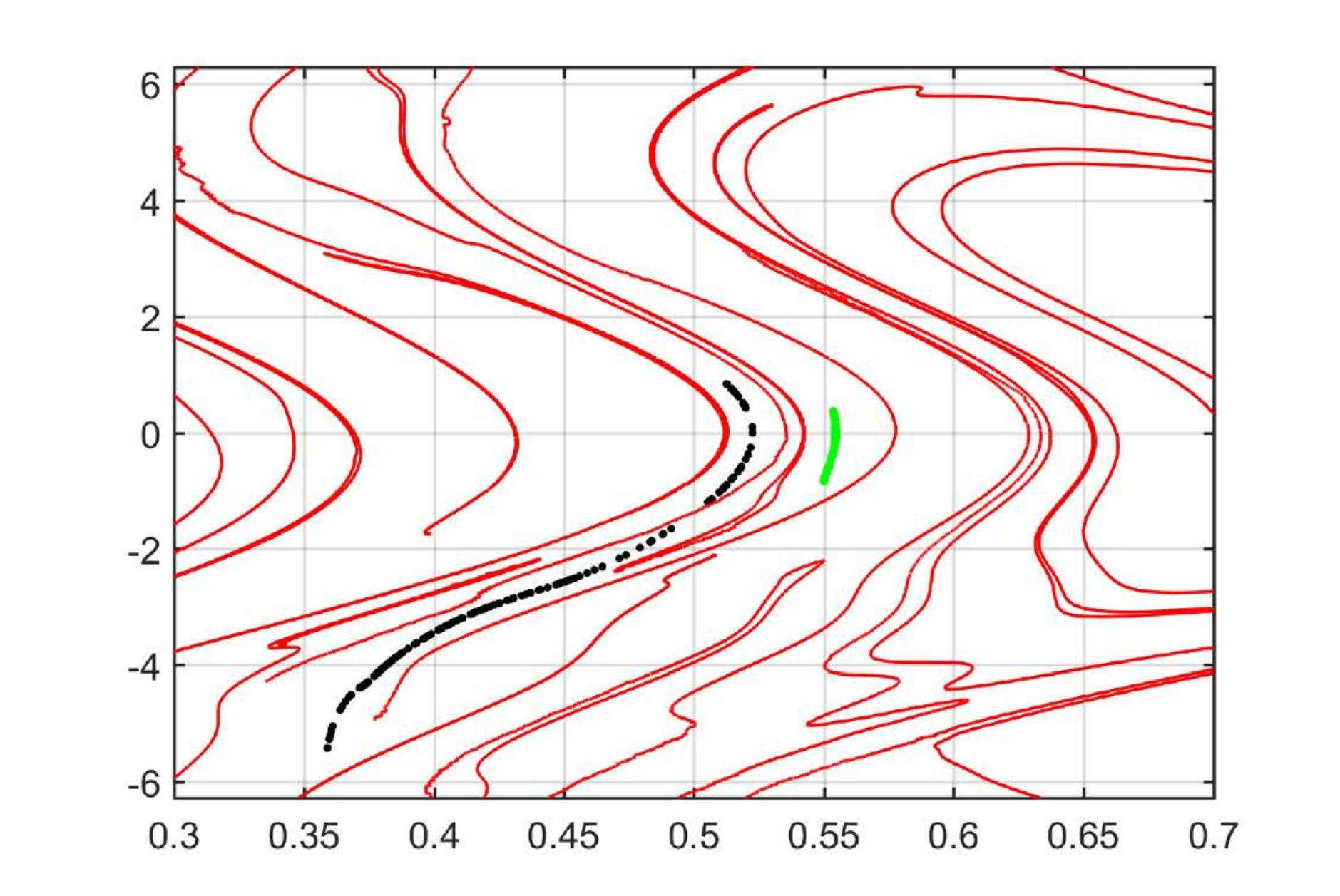}
\includegraphics[height=8cm,width=7.8cm]{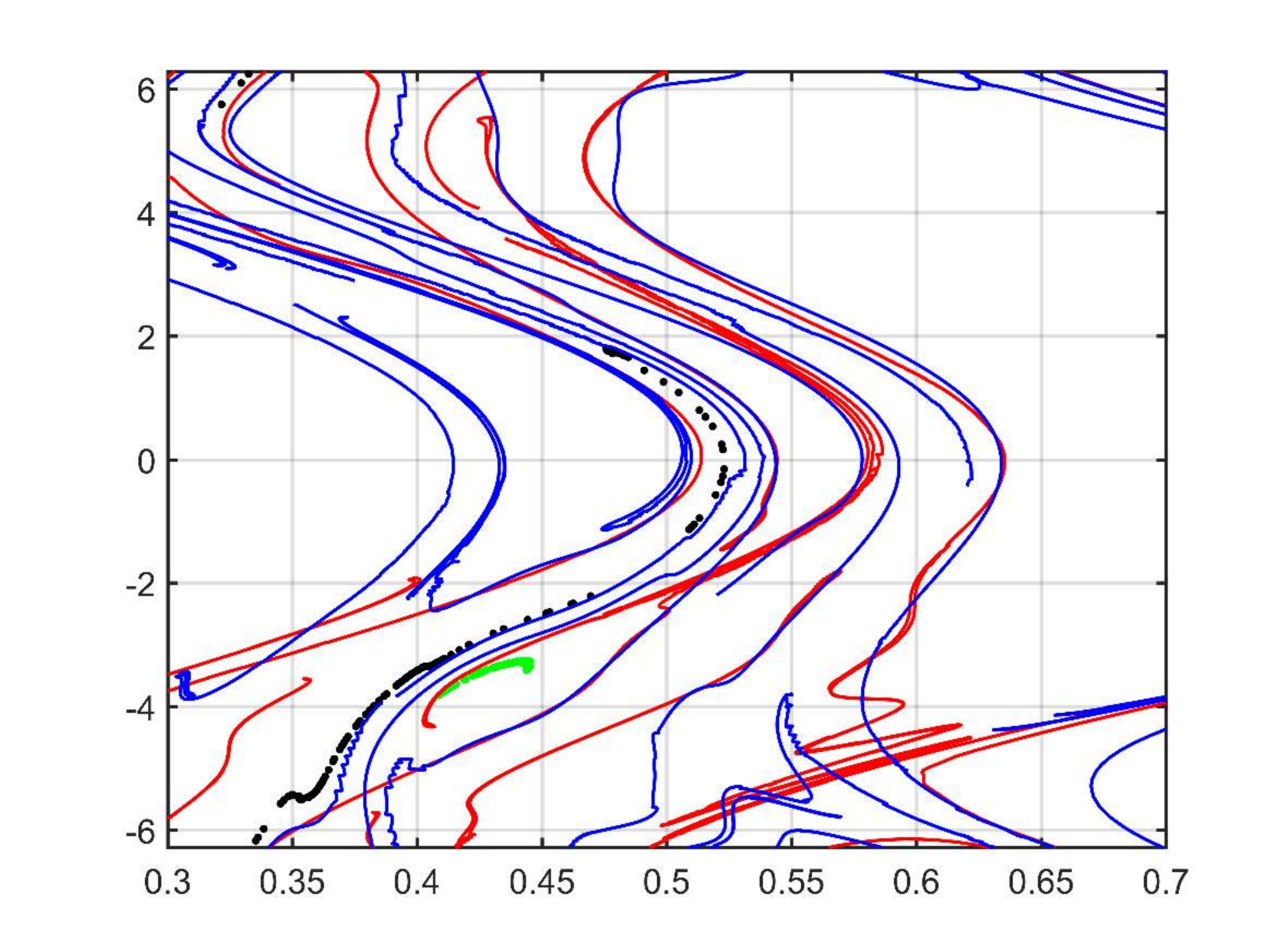}
\caption{Evolution of  the initial conditions and of  the repelling LCS of Fig. \ref{tempo_0} at $t=416.5$ (left panel) and $t=417$ (right panel). The LCS have been obtained with the system that evolves from $t=416.5$ to $t=418.5$ (left panel) and  from $t=417$ to $t=419$ (right panel). In the right panel, the attracting LCS (blue curves) have been obtained with the system that evolves from  $t=415$ to $t=417$. This is just to show how particles arrange themselves along attracting LCS computed for the same time interval of the particles evolution.}
\label{tempi_finali}
\end{figure*}

In Fig.\ref{tempo_0}  two sets, each with $75$ initial conditions, (marked green and black) are located on the two sides of a repelling LCS. The i.c. in each set are inside a circle  with radius   equal to 0.003. 
In Figs. \ref{tempi_iniziali}-\ref{tempi_finali} the evolution of these two sets is shown. {During the first part of the evolution, Fig. \ref{tempi_iniziali},  the particles move away from  the nearby repelling LCS positioning
themselves in such a way  as to maximize their stretching in the perpendicular direction with respect to the LCS. In this phase the two set of i.c. behave similarly}. In the left panel of Fig.\ref{tempi_intermedi} it  appears clearly   that after only $\Delta t = 0.5$  the two sets of conditions have evolved obeying two different kinds of dynamics. Few time intervals are sufficient to recognize the chaotic dynamics of the black initial conditions: their distribution becomes more  stretched and  {convoluted}  than that of the green conditions since they are influenced by the presence of a nearby attractive structure. This tendency is more and more evident  with increasing time,  as shown in Figs.\ref{tempi_intermedi}-\ref{tempi_finali}. 

\begin{figure*}[h!]
\centering
\includegraphics[height=8cm,width=8cm]{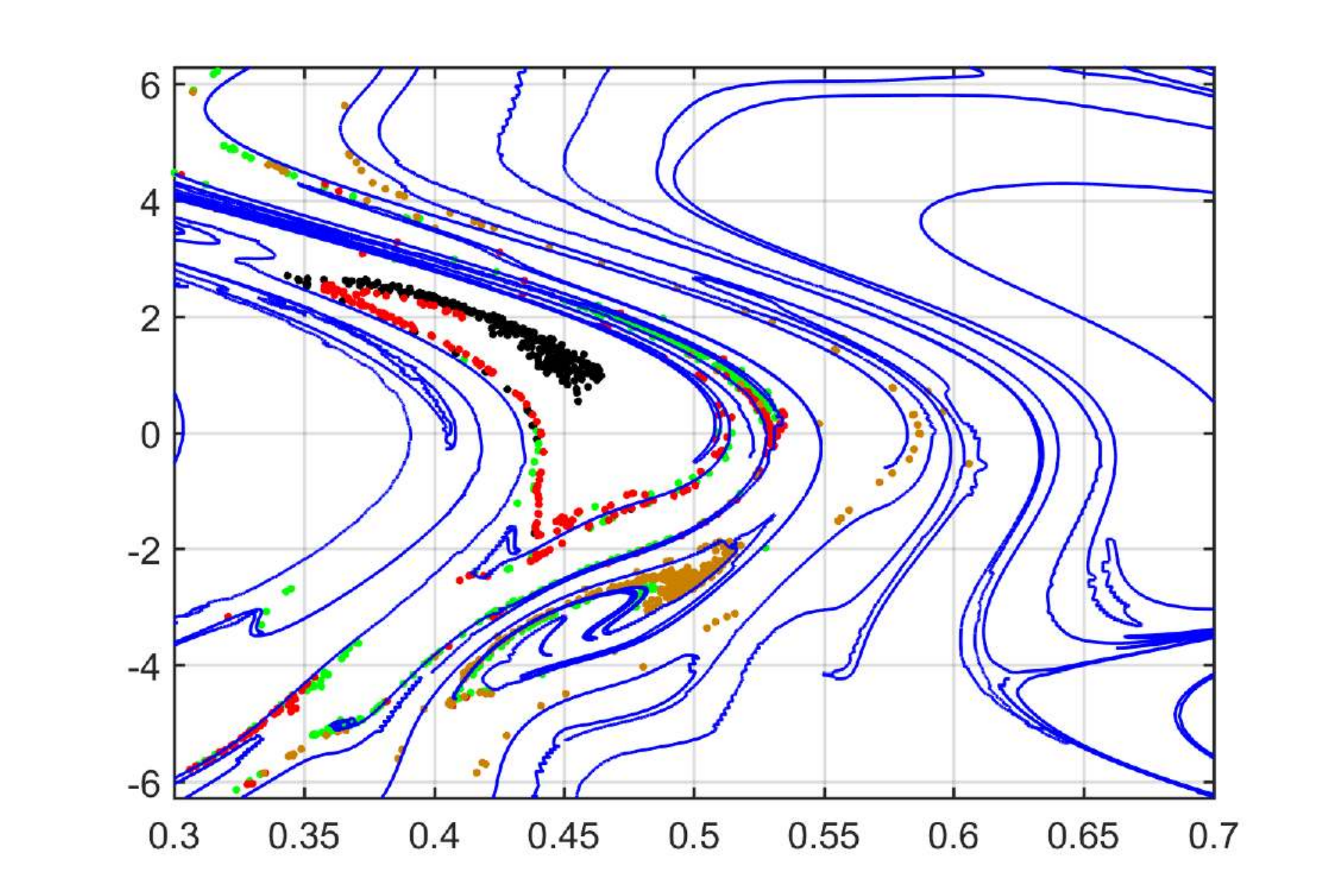}
\includegraphics[height=8cm,width=8cm]{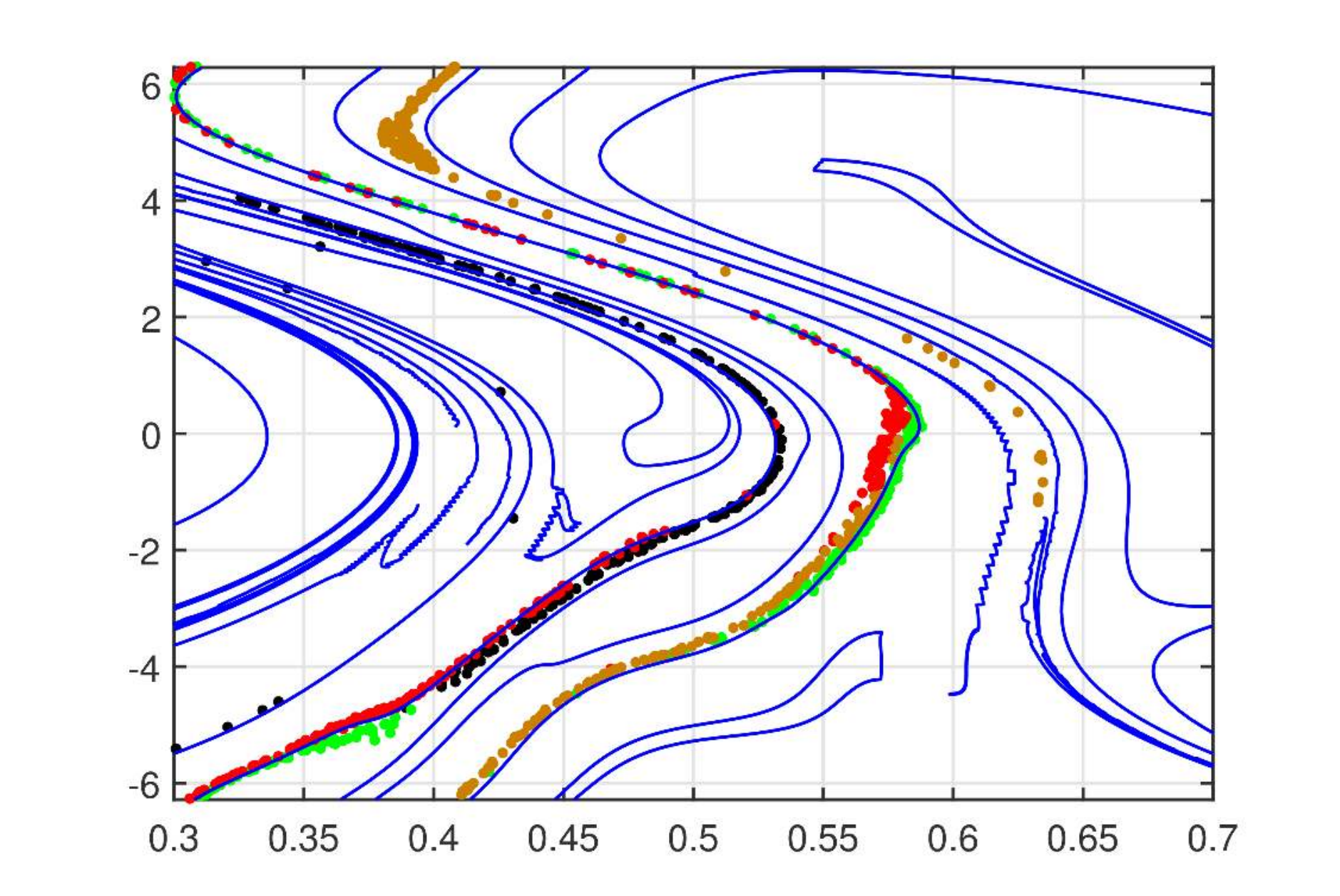}
\caption{Evolution of particles with $V=1000$ according to  the corresponding attractive LCS. For both figures the initial positions are the same: black particles $x=[0.5, 0.55], y=[0, -1]$, red $x=[0.5, 0.55], y=[-1, -2]$, green $x=[0.55, 0.6], y=[0, -1]$, brown $x=[0.55, 0.6], y=[-1, -2]$. There are $500$ i.c. for each color.
Particles in left panel start at time $t=416$ and the figure shows their position at $t=418$. Particles in the right panel start  at  $t=418$ and in the plot  their position at $t=419$ is shown.}
\label{square_v=1000}
\end{figure*}

\begin{figure*}[h!]
\centering
\includegraphics[height=8cm,width=7cm]{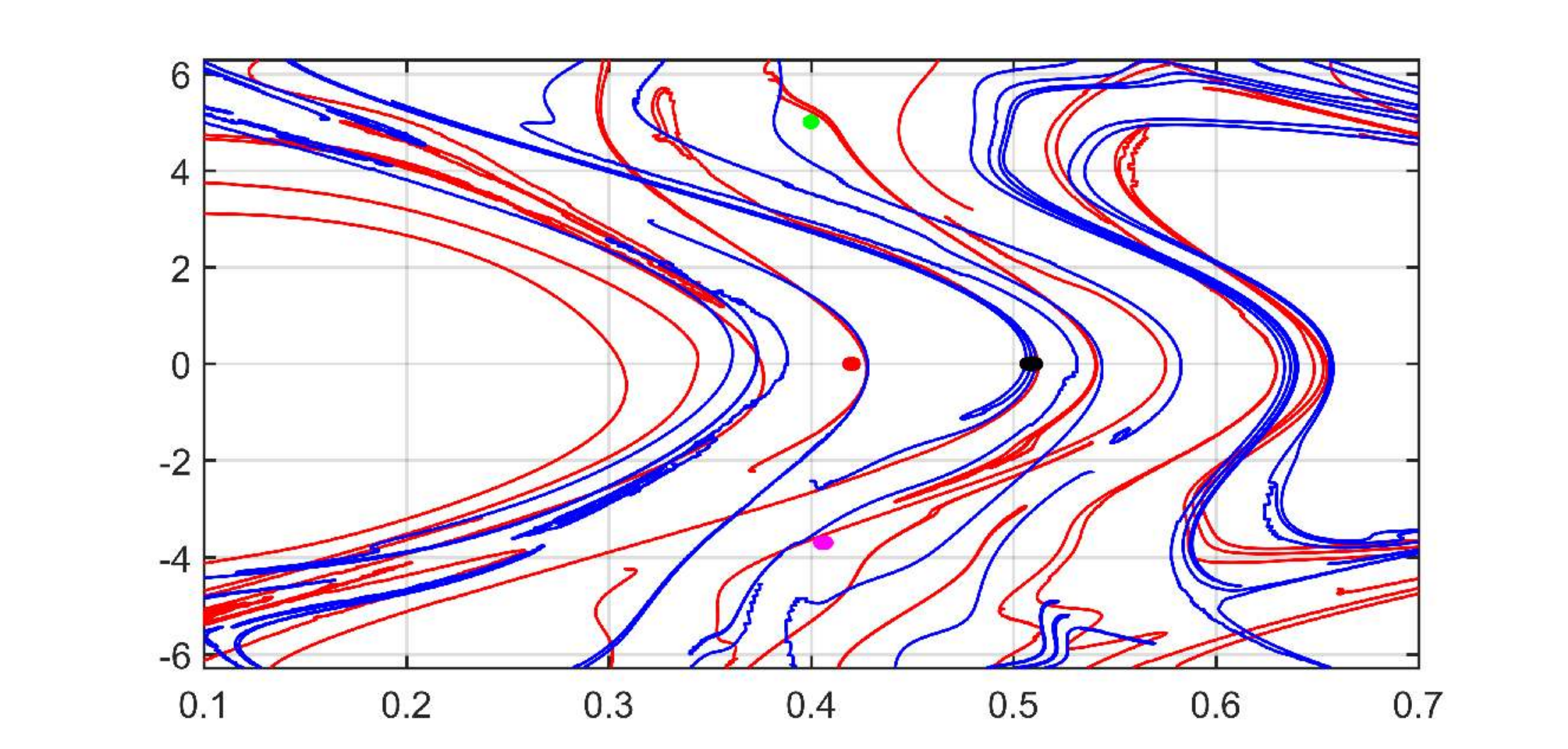}
\includegraphics[height=8cm,width=9cm]{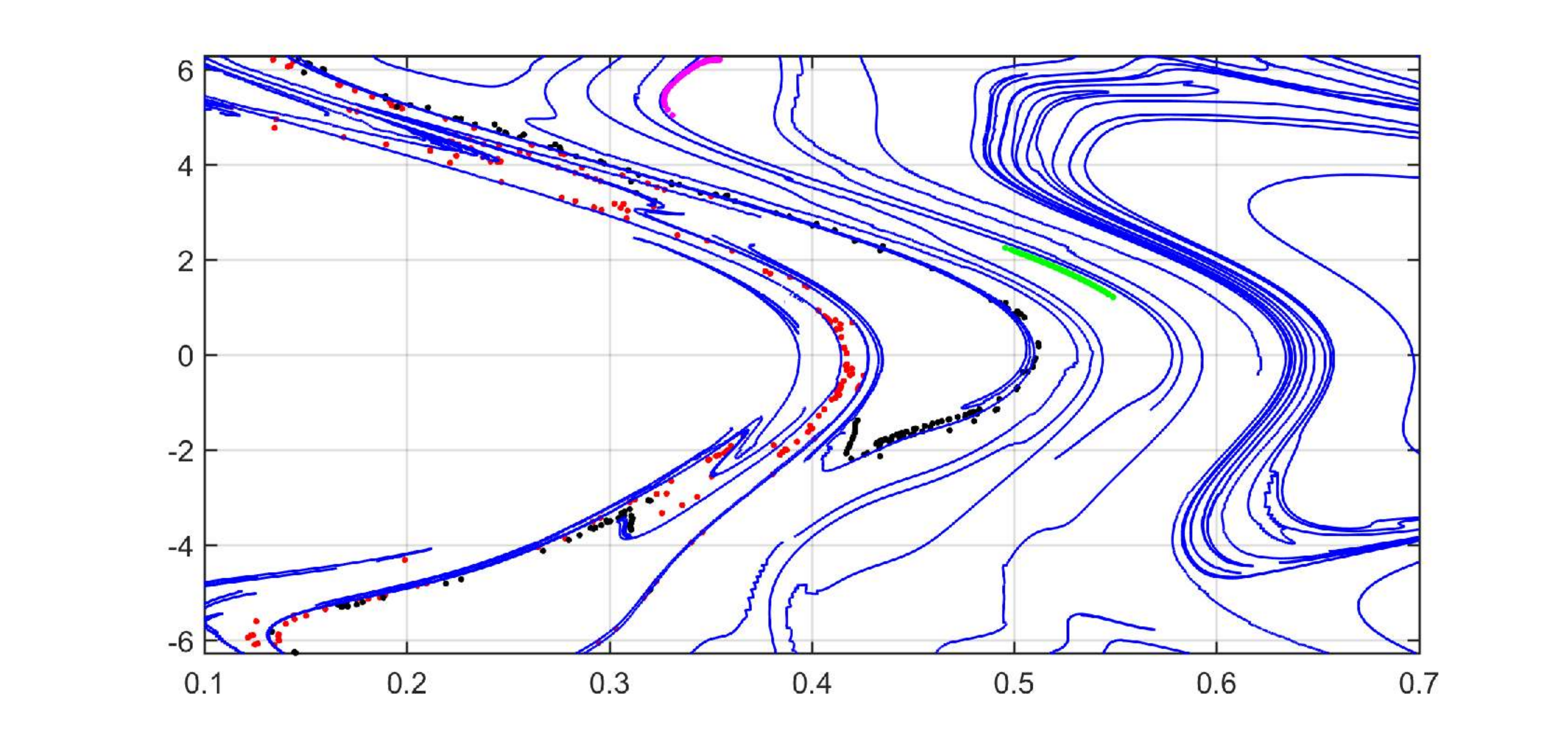}
\caption{Evolution of particles with $V= 1000$. On the left  frame the initial conditions at time $t=415$  are shown and on the right their  evolution at time $t=417$ is plotted. The aim of the picture is to underline how attractive LCS act as a  skeleton for  the dynamics. For clarity  we only show  attractive LCS (blue curves).
 Each  color  corresponds to  $300$ i.c. }
\label{4_ci}
\end{figure*}

These results  make the role of repelling and attractive LCS  evident when describing  the evolution of the system. In particular  the presence of a nearby attractive LCS seems to give rise to faster mixing phenomena. 
 In Figs.\ref{square_v=1000} and \ref{4_ci} we show  how  particles  feel the presence of attractive LCS.
In Fig. \ref{4_ci} we use spatially  localized initial conditions and  show that  these i.c. arrange themselves along the corresponding nearest attractive LCS. On the contrary, in Fig. \ref{square_v=1000} we use  spatially spread initial conditions covering the region $x = [0.5, 0.6]$, $y=[0, -2]$.

There are $500$ i.c. for each color: black particles in the region defined by $x=[0.5, 0.55]$ and $y=[0, -1]$; red particles in the region defined by $x=[0.5, 0.55]$ and $y=[-1, -2]$; green particles in the region defined by $x=[0.55, 0.6]$ and $y=[0, -1]$; brown particles in the region defined by $x=[0.55, 0.6]$ and  $y=[-1, -2]$.  The left panel shows  the positions of the  particles, starting at time $t=416$, at the time $t=418$. The blue lines are the attractive LCS computed starting from $t=418$ to $t=416$. 
The right panel shows  the positions of particles, starting at time $t=418$, at the time $t=419$. The blue lines are the attractive LCS computed starting from $t=419$ to $t=418$.\\
The interesting behavior is that although the i.c. cover a wide  region they evolve in such a way  as to position  themselves according to the attractive LCS. This  is the reason why often LCS are referred  to as the  "skeleton" of the dynamics.  This behavior is even more evident in the right panel of Fig. \ref{square_v=1000},  where we present the results obtained  for a shorter integration time interval ( $\Delta t =1$ ). In general, better results are obtained if shorter time intervals are used and this could be due to two reasons: the first  one is linked to the fact that  we use  a linear approximation in deriving the Cauchy-Green tensor, and the second one is  a consequence of the fact  that when the time interval increases the structures become much more convoluted  so that, in order to evaluate them in a suitable way, it is necessary to have a higher spatial resolution.

\begin{figure*}[h!]
\centering
\includegraphics[height=8cm,width=7cm]{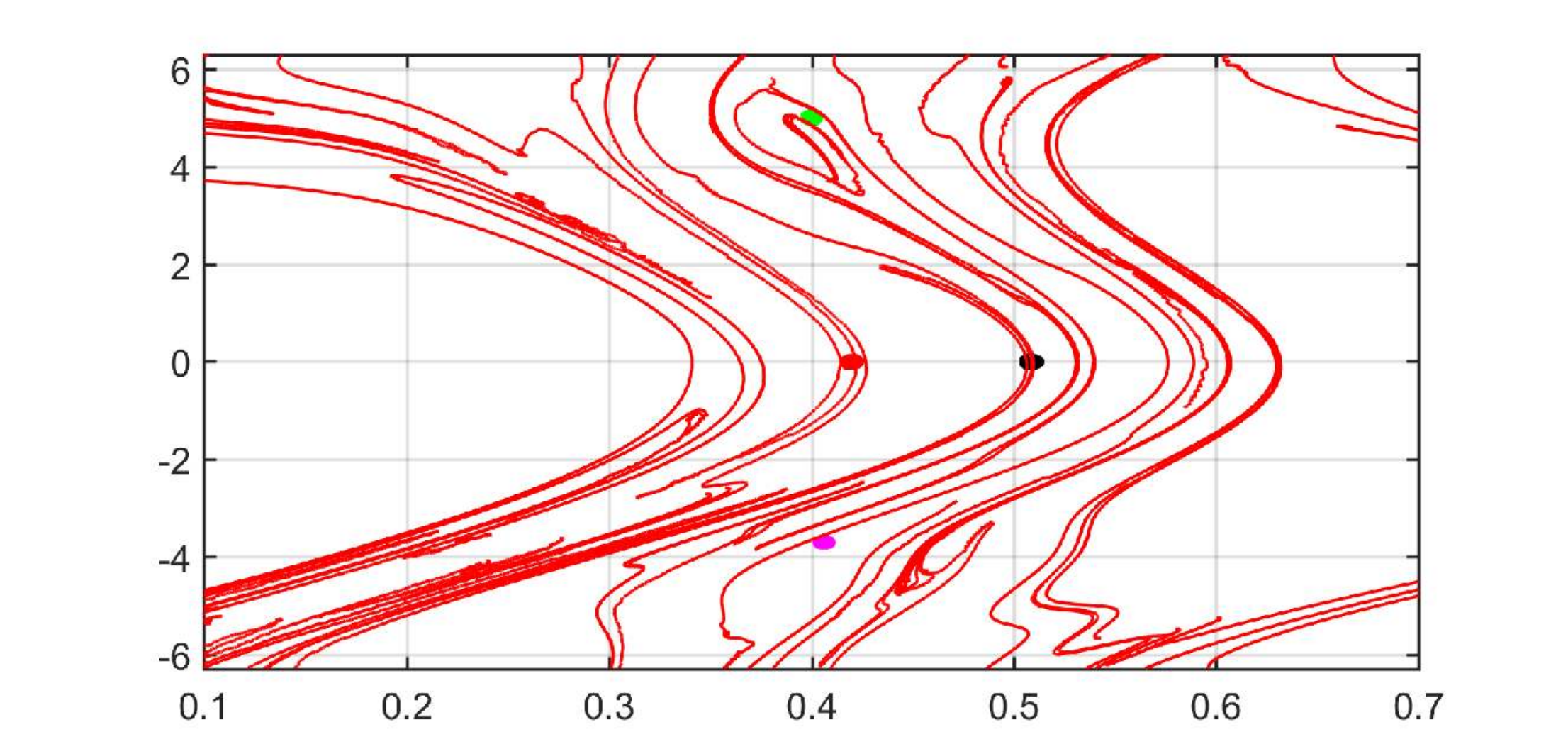}
\includegraphics[height=8cm,width=9cm]{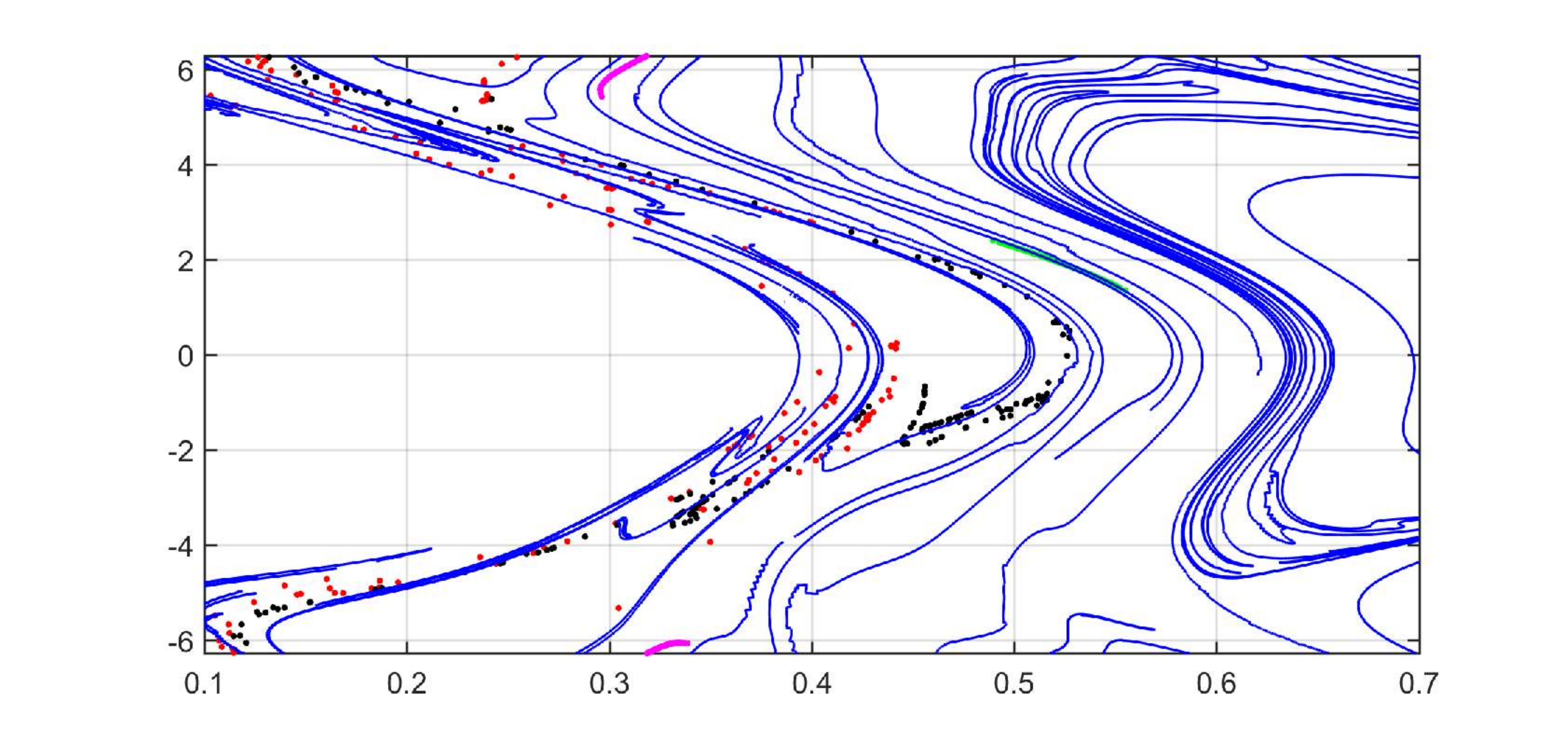}
\caption{ {Left panel: initial position at time $t=415$ of  particles and repelling LCS computed for particles with $V = 200$ with integration path $ \Delta z=19.9 $  z-loops (corresponding to $ \Delta t=10 $ ). Right panel: new position of  particles having $V = 200$ at time $t=425$ with attracting LCS computed for particles with $V = 1000$ with integration path $ \Delta z=19.9 $ z-loops (corresponding to $ \Delta t=2 $ ). }}
\label{4_ci_v_200}
\end{figure*}

Finally, we  compare  the LCS calculated for particles with ${V}=1000$ with the dynamics of particles with different velocities. Taking the same initial conditions depicted in Fig. \ref{4_ci} but this time with ${V}=200$, we find, {as shown in Fig. \ref{4_ci_v_200}}, that they behave differently with respect to the LCS {calculated for ${V}=1000$}. In the left panel of  Fig. \ref{4_ci_v_200}  we show  the initial position of the particles and the  repelling LCS computed integrating for  $ \Delta t=10 $ assuming ${ V}=200$  and, in the right panel, the new position of particles after $ \Delta t=10 $  with the attracting LCS computed with ${ V}=1000$ (to show how particles with $V=200$ can cross barriers obtained with velocity $V=1000$) and $ \Delta t=2 $. The choice  $ \Delta t=10 $ has been made  in order to have the same number of z-loops ( about 10 ) of the case ${ V}=1000$ shown in Fig. \ref{4_ci} and compare the results.  While in Fig. \ref{4_ci}   the particles arrange themselves along the attractive LCS,  in   Fig. \ref{4_ci_v_200} on the contrary  such  a relation between the attractive LCS (computed for particles with ${ V}=1000$) and the position of the particles (having ${ V}=200$) is not present. 
In particular, although the particle positions appear qualitatively similar to those in Fig. \ref{4_ci},  black particles with ${V}=200$ are shifted with respect to black particles having ${ V}=1000$. This is due to the fact that the magnetic field configuration at time $t=425$ (see Poincar\`e plot in Fig. \ref{PP2}) has the {  $m = 2 $ island chain}  shifted to the right with respect to its position at $t=415$ (see Poincar\`e  plot in Fig. \ref{PP1}). The same explanation also holds for some red particles that seem to be able to cross  through a region  that at time $t=415$ is regular.
Moreover, we can see also that red particles with $V=200$ behave "more chaotically" than red particles in Fig. \ref{4_ci}. This is probably due to the fact that,  as can be seen  in  the left panel of Fig. \ref{4_ci_v_200}, red i.c. are divided  by two  repelling LCS (red curves).
 Finally, also the magenta particles  find   themselves in a different region with  respect to the case ${ V}=1000$ (although their behavior  in the two cases 
 is quite similar).

\section{Conclusions}

Lagrangian Coherent Structures have been shown  to  provide  a very convenient tool in order to identify in a compact and easily visualizable way  the main features of the dynamics of the physical system under consideration. Clearly,  with large computers and long integration times one can recover all the needed  information just following the individual  trajectories of a large number of initial conditions. However LCS  do not simply provide the salient features that can be extracted from such large scale integrations but provide a  framework and a language  to be used in characterizing the  evolution of such features in time.

In this and in the accompanying paper (Article {\it I}) we have applied the LCS tool to the dynamics of charged particles in a magnetized plasma in the presence of a time evolving  reconnection instability.
The LCS  method is generally applicable  without approximations by referring to the full particle dynamics in $3D$ coordinate space and  by employing e.g.,  the exact  particle Hamiltonian in time varying electromagnetic fields.  Here, however, we have made use of two important simplifications with the aim of illustrating the method more than of obtaining exact results  to be applied to a specific fusion experiment configuration.
In the first  model  we have used the magnetic field lines in a slab configuration at a given physical instant of time  as a proxy for the particle trajectories. In the second model we have introduced an elementary  procedure  in order to account  in an approximate way for the fact that the magnetic configuration evolves in time during the particle motion.  In both cases we have been confronted with a $2D$ phase space,  much simpler  than the full $6D$ phase space that would be required  when solving the full particle dynamics. Clearly, this major simplification has been made possible by the fact that in the adopted configuration  a strong and almost uniform magnetic field is present that, in a toroidal laboratory configuration,  would correspond to the toroidal field.
The first model has allowed us to relate  the structures that govern the global dynamics of the particles to the evolution of the magnetic islands due to the development of magnetic reconnection. The second model has allowed us to show, even if  in a rather schematic way, that these structures  depend on the particle velocity (i.e. indirectly) on the particle energy.
We conclude by reiterating that the methods  developed in these two papers can be extended  to more refined  dynamical descriptions such as e.g  to a description based  on  the particle gyrokinetic approximation.

\section*{Acknowledgments}
GDG and DG thank Dario Borgogno for fruitful discussions.
Computational resources provided by hpc@polito, which is a project of Academic Computing within the Department of Control and Computer Engineering at the Politecnico di Torino (http://www.hpc.polito.it)

\bibliographystyle{unsrt}
\bibliography{bibliolcs}

\begin{thebibliography}{1}

\bibitem{haller2000lagrangian}
George Haller and Guocheng Yuan.
\newblock Lagrangian coherent structures and mixing in two-dimensional
  turbulence.
\newblock {\em Physica D: Nonlinear Phenomena}, 147(3):352--370, 2000.

\bibitem{borgogno2005aspects}
Dario Borgogno, Daniela Grasso, F~Porcelli, F~Califano, F~Pegoraro, and
  D~Farina.
\newblock Aspects of three-dimensional magnetic reconnection.
\newblock {\em Physics of plasmas}, 12(3):032309, 2005.

\bibitem{borgogno2008stable}
Dario Borgogno, D~Grasso, F~Pegoraro, and TJ~Schep.
\newblock Stable and unstable invariant manifolds in a partially chaotic
  magnetic configuration generated by nonlinear reconnection.
\newblock {\em Physics of Plasmas (1994-present)}, 15(10):102308, 2008.

\bibitem{borgogno2011barriers}
Dario Borgogno, Daniela Grasso, F~Pegoraro, and TJ~Schep.
\newblock Barriers in the transition to global chaos in collisionless magnetic
  reconnection. i. ridges of the finite time lyapunov exponent field.
\newblock {\em Physics of Plasmas (1994-present)}, 18(10):102307, 2011.

\bibitem{onu2015lcs}
K~Onu, Florian Huhn, and George Haller.
\newblock Lcs tool: A computational platform for lagrangian coherent
  structures.
\newblock {\em Journal of Computational Science}, 7:26--36, 2015.

\end{thebibliography}

\end{document}